\documentclass[12pt,a4paper]{article}

\usepackage{datetime} \usdate
\usepackage[utf8]{inputenc}
\usepackage[T1]{fontenc}
\usepackage{amsmath,amssymb,amsfonts,amsthm}
\usepackage{mathtools}
\usepackage{bm} 
\usepackage[margin=1in]{geometry}
\usepackage{graphicx}
\usepackage{multirow}
\usepackage{booktabs}     
\usepackage{tabularx}  
\usepackage{natbib}   
\usepackage[hidelinks]{hyperref}
\usepackage{setspace}
\usepackage{float}    
\usepackage{caption}
\usepackage{subcaption}
\usepackage{xcolor}
\usepackage{enumitem}
\usepackage{tikz}
\usetikzlibrary{patterns}
\usepackage{blkarray}  % for annotated block matrices
   
\newcommand{\boxsize}{12pt}
\newcommand{\vlbox}{\tikz[baseline=-0.5ex] \node[draw, minimum width=\boxsize, minimum height=\boxsize, pattern=vertical lines, pattern color=black!50, rounded corners] {};}
\newcommand{\dlbox}{\tikz[baseline=-0.5ex] \node[draw, minimum width=\boxsize, minimum height=\boxsize, pattern=north east lines, pattern color=black!50, rounded corners] {};}
\newcommand{\hlbox}{\tikz[baseline=-0.5ex] \node[draw, minimum width=\boxsize, minimum height=\boxsize, pattern=horizontal lines, pattern color=black!50, rounded corners] {};}
\newcommand{\dtbox}{\tikz[baseline=-0.5ex] \node[draw, minimum width=\boxsize, minimum height=\boxsize, pattern=dots, pattern color=black!50, rounded corners] {};}

\title{{\textbf{Scalable Clustered Network Connectedness \\with Control Variables:\\ Theory and Application to Global Banking}}}

\author{
	Bastien Buchwalter\thanks{SKEMA Business School and Universit\'{e} C\^{o}te d'Azur. Email: \texttt{bastien.buchwalter@skema.edu}}
	\and
	Francis X.\ Diebold\thanks{University of Pennsylvania and NBER. Email: \texttt{fdiebold@sas.upenn.edu}}
	\and
	Kamil Y{\i}lmaz\thanks{Ko\c{c} University. Email: \texttt{kyilmaz@ku.edu.tr}}
}
 
\bigskip

\date{\normalsize \vspace{4mm} First Draft: October 2025\\ This Draft: \today}

\begin{document}
	
	\maketitle
	\thispagestyle{empty}

	\noindent \textbf{Abstract}: We extend the clustered connectedness framework of \citet{BuchwalterDieboldYilmaz2026} in two complementary directions that improve the robustness and interpretability of cross-cluster connectedness. First, we develop a diagnostic for residual ordering sensitivity by characterizing the distribution of cluster-level net connectedness across all admissible identification orderings and, in particular, by pairing first- and last-position orderings while holding fixed the relative ordering of all other clusters. Second, we introduce a dedicated cluster of control variables to absorb variation associated with observed common macro-financial factors while preserving the computational scalability of the clustered framework. The control cluster is fixed first, and bank innovations are residualized with respect to it before the remaining bank clusters are permuted and orthogonalized as usual, leaving the number of admissible bank-cluster identification strategies unchanged. Under the maintained recursive assumption that control-cluster innovations are contemporaneously exogenous to bank-cluster innovations, the remaining cross-cluster connectedness among the bank clusters can be interpreted as bank-to-bank transmission net of those observed common-factor shocks. We apply the methodology to seventy-one global banks grouped into seven regional clusters over 2003--2024. The treatment of common macro-financial factors materially affects both system-wide cross-group connectedness and cluster-level net positions. Placing the controls in a dedicated first cluster also substantially reduces paired first-versus-last ordering sensitivity across all seven bank clusters, with especially large reductions for the United States and the European clusters.
	
	\bigskip
	
	\bigskip
	
	\noindent\textbf{Acknowledgments:} We gratefully acknowledge useful input from seminar and conference participants. 
	
	
	
%	\newpage
%	\pagestyle{empty}
%	\tableofcontents
%	
	
	\newpage
	
	\onehalfspacing
	
	\pagestyle{plain}
	\setcounter{page}{1}

	\section{Introduction}   \label{sec:intro}
	\thispagestyle{empty}
	
	Network connectedness has emerged as a central empirical object in a wide range of economic environments, from financial markets to macroeconomic aggregates and international linkages of various kinds \citep{DY2023}. The Diebold--Y{\i}lmaz (DY) framework \citep{DieboldYilmaz2009} provides the foundation for a large and growing literature that measures how shocks propagate through economic and financial systems. Recently, \citet{BuchwalterDieboldYilmaz2026} extended the DY framework by introducing clustered identification, in which shocks are grouped into economically meaningful blocks such as asset classes, industries, or regions. Shocks are assumed orthogonal across clusters while remaining correlated within clusters, yielding a useful middle ground between fully orthogonalized and fully generalized identification strategies. We refer to this clustered-identification approach as the BDY framework.
	
	The BDY framework greatly reduces the dimensionality of identification relative to traditional orthogonalized connectedness analysis. Nevertheless, cluster ordering remains relevant because connectedness measures continue to depend on the sequence in which clusters enter the identification strategy. Although the number of admissible orderings is substantially smaller than in the original DY setting, ordering may still materially influence empirical conclusions. Understanding and quantifying this residual ordering sensitivity is therefore important, particularly when connectedness measures are given a structural or causal interpretation conditional on the chosen identification.
	
	This paper develops the BDY framework in two complementary directions. First, we develop a diagnostic for residual ordering sensitivity. We characterize the distribution of each cluster's net connectedness across all admissible cluster orderings and then isolate position effects using paired comparisons. For a given cluster, we compare orderings in which it is placed first and last while holding fixed the relative ordering of all other clusters, and we examine the resulting distribution of paired changes in net connectedness. In the absence of ordering effects, that distribution should concentrate near zero. Empirically, however, the paired differences are substantial for several clusters, indicating that recursive ordering materially affects measured transmission. The diagnostic therefore provides a direct way to distinguish the level of measured connectedness from its robustness to identification.
	
	Second, we introduce a dedicated cluster of control variables. The key issue is not simply whether controls are included, but how they enter the identification. Placing controls within the existing regional clusters blurs the distinction between innovations associated with the controls and innovations originating in the banks, so control-related common variation can remain embedded in measured bank connectedness. Placing the controls in a separate cluster preserves that distinction, but treating the enlarged set of clusters symmetrically would increase the number of admissible orderings and weaken the scalability that motivates clustered identification. We resolve both issues by fixing the control cluster first, residualizing the bank innovations with respect to it---equivalently, replacing the bank-innovation covariance matrix by the corresponding Schur complement---and then applying the usual BDY block orthogonalization across all admissible orderings of the bank clusters. The resulting specification preserves the computational merits of the BDY framework while substantially reducing ordering sensitivity.
	
	The control-first specification also sharpens the interpretation of cross-cluster connectedness. Because the bank innovations are residualized with respect to the control cluster before the bank clusters are orthogonalized, contemporaneous linear variation associated with the observed common macro-financial controls is attributed to the control cluster first. Under the maintained recursive identification in which innovations in the control cluster are contemporaneously exogenous to innovations in the bank clusters, the remaining cross-cluster connectedness among the bank clusters can be interpreted as bank-to-bank transmission net of those observed common-factor shocks. Without imposing that structural interpretation, the specification still provides a transparent descriptive decomposition: bank-cluster connectedness after partialing out contemporaneous linear variation associated with the observed controls.
	
	Taken together, these contributions combine an explicit diagnostic for residual ordering sensitivity with a scalable control-variable architecture. The resulting framework improves the robustness and interpretability of clustered connectedness measures while preserving computational viability.
	
	The remainder of this paper is structured as follows. In Section~\ref{sec:framework}, we develop the measurement and identification framework for clustered connectedness, including the distinction between system-wide and cluster-level connectedness measures, the ordering-sensitivity diagnostic, and the control-variable extension. Section~\ref{sec:results} applies the framework to global banking: we describe the data and empirical implementation, present connectedness results under alternative treatments of the controls, evaluate ordering sensitivity, and synthesize the economic interpretation across identification strategies. Section~\ref{sec:conclusion} concludes.

	\section{Measuring and Identifying Clustered Network Connectedness}\label{sec:framework}
	
	This section develops the measurement and identification framework used throughout the paper. We first review the BDY clustered network connectedness framework, which provides a computationally scalable approach to measuring connectedness in large networks. We then examine the residual ordering sensitivity that remains under clustered identification, develop a diagnostic for measuring it, and introduce our control-variable extension. The extension incorporates control variables into the identification strategy to substantially reduce ordering sensitivity while maintaining the computational advantages of the clustered framework. This methodology serves as the basis for the subsequent empirical analysis.
	
	\subsection{VAR and clustered connectedness}
	
	We begin with an $N$-variable, covariance-stationary VAR($p$) of the form $\mathbf{x}_t=\sum_{i=1}^{p}\boldsymbol{\Phi}_i \mathbf{x}_{t-i}+\mathbf{u}_t,$
	where $\mathbb{E}[\mathbf{u}_t]=0$ and $\mathbb{V}[\mathbf{u}_t]=\boldsymbol{\Sigma}$ for all $t$. Under covariance stationarity, the system admits a moving-average representation,
	$\mathbf{x}_t=\sum_{i=0}^{\infty}\boldsymbol{A}_i\,\mathbf{u}_{t-i}$,
	where $\boldsymbol{A}_0 = I_N$ and, for $i \ge 1$, the coefficient matrices satisfy the recursion $\boldsymbol{A}_i=\boldsymbol{\Phi}_1\boldsymbol{A}_{i-1}+\cdots+\boldsymbol{\Phi}_p\boldsymbol{A}_{i-p}$,
	with $\boldsymbol{A}_i=\mathbf{0}$ for $i<0$. Because the reduced-form residuals $\mathbf{u}_t$ are typically correlated, it is convenient to transform them into shocks that are orthogonal across clusters. Let $\boldsymbol{Q}_C$ be any nonsingular block lower-triangular matrix, and define $\boldsymbol{\epsilon}_t = \boldsymbol{Q}_C^{-1}\mathbf{u}_t$, where $C$ denotes the number of clusters. It follows that $\mathbb{E}[\boldsymbol{\epsilon}_t]=0$, $\mathbb{V}[\boldsymbol{\epsilon}_t]=\boldsymbol{\Omega}_C$,
	and the moving-average representation can be rewritten as
	\begin{align*}
		\mathbf{x}_t=\sum_{i=0}^{\infty}(\boldsymbol{A}_i \boldsymbol{Q}_C)\,\boldsymbol{\epsilon}_{t-i}.
	\end{align*}
	The matrix $\boldsymbol{Q}_C$ provides a decomposition of the system's dynamics into within-cluster co-movement and cross-cluster transmission. Such a structure is especially useful in settings with a large number of assets when shocks are to be orthogonalized across economically meaningful clusters. Under maintained recursive identifying assumptions, the resulting cross-cluster shocks can also be given a causal interpretation.
	
	To study the dynamic effect of shocks, consider an isolated, one-period disturbance to the $j$th orthogonalized innovation $\epsilon_{j,t}$. Using the linearity of conditional expectations, the $h$-step-ahead response of all variables is given by
	\[
	\boldsymbol{\psi}_j^{C}(h)
	= (\boldsymbol{A}_h \boldsymbol{Q}_C)\,\mathbb{E}[\boldsymbol{\epsilon}_t \mid \epsilon_{j,t}=\delta_j]
	= \frac{(\boldsymbol{A}_h \boldsymbol{Q}_C)\,\boldsymbol{\Omega}_C \mathbf{e}_j \,\delta_j}{\omega_{C,jj}},
	\]
	where $\mathbf{e}_j$ denotes the $j$th selection vector and $\omega_{C,jj}$ is the variance of $\epsilon_{j,t}$. By choosing $\delta_j=\sqrt{\omega_{C,jj}}$, we obtain the impulse responses to a one-standard-deviation orthogonalized shock:
	\[
	\boldsymbol{\psi}_j^{C}(h)
	= \frac{\boldsymbol{A}_h \boldsymbol{Q}_C\,\boldsymbol{\Omega}_C \mathbf{e}_j}{\sqrt{\omega_{C,jj}}}. \label{eq:irf}
	\]
	
	The orthogonalized moving-average representation also provides the basis for variance decompositions (VDs), which allocate the $H$-step-ahead forecast error variance of each component $\mathbf{x}_{i,t}$ to each shock $\epsilon_{j,t}$. That is, variance decompositions quantify the fraction of the forecast error variance of variable $i$ that is attributable to shock $j$, thereby revealing the channels through which innovations propagate throughout the system. We denote the $H$-step-ahead VD by \(\tilde{\theta}^{C}_{ij}(H)\):
	\begin{align}
		\tilde{\theta}^{C}_{ij}(H) = \frac{\sum_{h=0}^{H-1} \left( \boldsymbol{e}_i' \boldsymbol{\psi}_j^C(h)\right)^2}{\sum_{h=0}^{H-1}(\boldsymbol{e}_i' \boldsymbol{A}_h \boldsymbol{\Sigma} \boldsymbol{A}_h'  \boldsymbol{e}_i)} = \frac{\omega_{C,jj}^{-1}\sum_{h=0}^{H-1} (\boldsymbol{e}_i' \boldsymbol{A}_h\boldsymbol{Q}_C\boldsymbol{\Omega}_C \boldsymbol{e}_j)^2}{\sum_{h=0}^{H-1}(\boldsymbol{e}_i' \boldsymbol{A}_h \boldsymbol{\Sigma} \boldsymbol{A}_h' \boldsymbol{e}_i)}, \label{thetasec}
	\end{align}
	where \(\tilde{\theta}_{ij}^C\) is the share of the $H$-step-ahead forecast error variance of asset $i$ due to shocks from asset $j$. We note that generally \(\sum_{j=1}^N\tilde{\theta}_{ij}^C(H)\neq1\). This stems from the non-zero off-diagonal elements of the covariance matrix. Following the spirit of \cite{DieboldYilmaz2012}, we normalize $\theta_{ij}^C(H) = \frac{\tilde{\theta}_{ij}^C(H)}{ \sum_{j=1}^N\tilde{\theta}_{ij}^C(H)}$ so that \(\sum_{j=1}^N\theta_{ij}^C(H)=1\).
	
	Having established the general clustered framework, we now examine two benchmark identification strategies corresponding to opposite extremes of the clustering problem. These corner solutions clarify the relationship among identification, ordering sensitivity, and computational viability, thereby motivating the control-variable extension developed later in the paper.
	
	\subsection{Corner solutions}
	
	The BDY framework nests the two identification strategies that dominate the connectedness literature as limiting cases. At one extreme, all shocks are orthogonalized. This yields a recursively identified system that can support a causal interpretation under the maintained ordering assumptions, but at the cost of severe ordering sensitivity. At the other extreme, no shocks are orthogonalized. This eliminates ordering concerns entirely but does not separately identify contemporaneous shocks and therefore does not by itself support a causal interpretation. The BDY framework occupies the intermediate region between these extremes by orthogonalizing shocks across clusters while preserving correlation within clusters.
	
	\subsubsection{Orthogonalized IRFs: $C = N$}
	
	The approach of \citet{Sims1980} sets $C = N$, so that each variable constitutes its own cluster. Here $\bm{Q}_N = \bm{M}$, where $\bm{M}$ is the unique lower-triangular Cholesky factor of $\bm{\Sigma} = \bm{M}\bm{M}^\top$, and $\bm{\Omega}_N = \bm{I}_N$. The orthogonalized IRFs are:
	\begin{equation*}
		\bm{\psi}_j^N(h) = \bm{\psi}_j^o(h) = \bm{A}_h \bm{M} \bm{e}_j.
		\label{eq:irf_ortho}
	\end{equation*}
	The resulting $\bm{\Omega}_N$ is an identity matrix, i.e., a block-diagonal structure with $N$ scalar blocks:
	\begin{equation*}
		\bm{\Omega}_N = \begin{pmatrix} \bm{I} & \bm{0} & \bm{0} \\ \bm{0} & \bm{I} & \bm{0} \\ \bm{0} & \bm{0} & \bm{I} \end{pmatrix}.
	\end{equation*}
	
	Under the maintained recursive ordering assumptions, this approach permits causal interpretation but requires selecting one of $N!$ possible orderings---a prohibitive combinatorial burden for even moderately sized systems.
	
	\subsubsection{Generalized IRFs: $C = 1$}
	
	The approach of \citet{KoopPesaranPotter1996} sets $C = 1$, grouping all variables into a single cluster. Here $\bm{Q}_1 = \bm{I}_N$ and $\bm{\Omega}_1 = \bm{\Sigma}$. The generalized IRFs are:
	\begin{equation*}
		\bm{\psi}_j^1(h) = \bm{\psi}_j^g(h) = \frac{\bm{A}_h \bm{\Sigma} \bm{e}_j}{\sqrt{\sigma_{jj}}}.
		\label{eq:irf_gen}
	\end{equation*}
	The resulting $\bm{\Omega}_1 = \bm{\Sigma}$ retains all cross-correlations:
	\begin{equation*}
		\bm{\Omega}_1 = \begin{pmatrix} \bm{\Sigma}_{11} & \bm{\Sigma}_{12} & \bm{\Sigma}_{13} \\ \bm{\Sigma}_{21} & \bm{\Sigma}_{22} & \bm{\Sigma}_{23} \\ \bm{\Sigma}_{31} & \bm{\Sigma}_{32} & \bm{\Sigma}_{33} \end{pmatrix}.
	\end{equation*}
	
	The comparison of these corner solutions highlights the central trade-off in connectedness analysis. Orthogonalized identification can support causal interpretation under maintained recursive identifying assumptions, but requires an ordering that allocates contemporaneous correlation across shocks. Generalized identification avoids ordering altogether but leaves contemporaneous correlation embedded in the connectedness measures and does not separately identify structural shocks. The BDY framework provides an intermediate solution by reducing the identification problem from $N!$ to $C!$ orderings while allowing causal interpretation of cross-cluster transmission under the maintained clustered recursive identification.
	
	The present paper addresses a remaining limitation of this intermediate solution. Although clustering substantially reduces the dimensionality of the ordering problem, common shocks may still be redistributed across clusters according to the chosen ordering. Later in this section, we show how this ordering sensitivity can be diagnosed and how a dedicated control cluster can substantially mitigate it without sacrificing the scalability of the BDY framework.
	
	The two corner solutions illustrate the extremes of the identification problem. We now return to the general clustered framework and introduce the connectedness measures that serve as the principal quantities analyzed throughout the remainder of the paper.
	
	\subsection{Connectedness Measures}
\label{subsec:connectedness_measures}

The normalized node-level variance shares in equation~\ref{thetasec} support two related but distinct aggregations that we use throughout the paper: system-wide connectedness measures and cluster-level connectedness measures. Keeping the two separate is useful because the economically meaningful groups used to summarize a variance decomposition need not coincide with the blocks used for shock identification. Let $r(i)\in\{1,\ldots,G\}$ denote the reporting group assigned to node $i$. When the reporting groups coincide with the clustered identification blocks, $G=C$; under generalized identification, by contrast, there is a single identification block even though the resulting variance decomposition can still be summarized by economically meaningful reporting groups.

At the system level, we define total connectedness as
\begin{equation*}
    \mathcal{C}^{\text{total}}
    =
    \frac{1}{N}
    \sum_{i=1}^{N}
    \sum_{\substack{j=1\\j\neq i}}^{N}
    \theta_{ij}.
\end{equation*}
We decompose total connectedness into within-group and cross-group components,
\begin{align*}
    \mathcal{C}^{\text{within}}
    &=
    \frac{1}{N}
    \sum_{i=1}^{N}
    \sum_{\substack{j=1\\j\neq i,\ r(j)=r(i)}}^{N}
    \theta_{ij},\\
    \mathcal{C}^{\text{cross}}
    &=
    \frac{1}{N}
    \sum_{i=1}^{N}
    \sum_{\substack{j=1\\r(j)\neq r(i)}}^{N}
    \theta_{ij},
\end{align*}
so that
\begin{equation*}
    \mathcal{C}^{\text{total}}
    =
    \mathcal{C}^{\text{within}}
    +
    \mathcal{C}^{\text{cross}}.
\end{equation*}
These are system-wide averages over nodes. The within/cross classification is descriptive unless it is coupled with an identification strategy. In particular, under generalized identification $\mathcal{C}^{\text{cross}}$ simply collects variance shares linking nodes in different reporting groups; under clustered recursive identification, when the reporting groups coincide with the identification blocks, the corresponding cross-cluster component is based on shocks that are orthogonalized across clusters and can be interpreted as transmission under the maintained identifying assumptions.

We next define cluster-level summaries. For a reporting cluster $c$ containing $N_c$ nodes, the average own-node variance share is
\begin{equation*}
    \Theta_c^{\text{own}}=
    \frac{1}{N_c}
    \sum_{i\in c}
    \theta_{ii}.
\end{equation*}
The within-cluster component is
\begin{equation*}
    \Theta_c^{\text{within}}=
    \frac{1}{N_c}
    \sum_{\substack{i,j\in c\\i\neq j}}
    \theta_{ij},
\end{equation*}
which measures the average share of forecast-error variance attributable to shocks originating from other nodes in the same cluster. It is useful to combine these two pieces into the own-cluster share,
\begin{equation*}
    \Theta_c^{\text{own cluster}}
    =
    \Theta_c^{\text{own}}
    +
    \Theta_c^{\text{within}}
    =
    \frac{1}{N_c}
    \sum_{i\in c}
    \sum_{j\in c}
    \theta_{ij}.
\end{equation*}
The cross-cluster connectedness transmitted from cluster $k$ to cluster $c$ is
\begin{equation*}
    \Theta_{c\leftarrow k}^{\text{cross}}=
    \frac{1}{N_c}
    \sum_{i\in c}
    \sum_{j\in k}
    \theta_{ij}.
\end{equation*}

Aggregating across all sources of variance yields
\begin{equation*}
    \Theta_c^{\text{own}}
    +
    \Theta_c^{\text{within}}
    +
    \Theta_{c\leftarrow\bullet}^{\text{cross}}=
    1,
\end{equation*}
or, equivalently,
\begin{equation*}
    \Theta_c^{\text{own cluster}}
    +
    \Theta_{c\leftarrow\bullet}^{\text{cross}}=
    1,
\end{equation*}
where
\begin{equation*}
    \Theta_{c\leftarrow\bullet}^{\text{cross}}=
    \sum_{k\neq c}
    \Theta_{c\leftarrow k}^{\text{cross}}
\end{equation*}
denotes the total cross-cluster connectedness received by cluster $c$. Similarly,
\begin{equation*}
    \Theta_{\bullet\leftarrow c}^{\text{cross}}=
    \sum_{k\neq c}
    \Theta_{k\leftarrow c}^{\text{cross}}
\end{equation*}
denotes the total cross-cluster connectedness transmitted by cluster $c$. To summarize the role of a cluster within the network, we define net connectedness as
\begin{equation}
    Net_c= \Theta_{\bullet\leftarrow c}^{\text{cross}}-
    \Theta_{c\leftarrow\bullet}^{\text{cross}}.\label{eq:net}
\end{equation}
Net connectedness therefore measures whether a cluster is a net transmitter or net receiver of cross-cluster shocks. That is, a positive value of $Net_c$ indicates that cluster $c$ is a net transmitter of shocks, while a negative value indicates that it is a net receiver.

The system-wide and cluster-level measures are linked by cluster-size weighting. If the reporting partition contains $G$ clusters and $N=\sum_{c=1}^{G}N_c$, then
\begin{align*}
    \mathcal{C}^{\text{within}}
    &=
    \sum_{c=1}^{G}
    \frac{N_c}{N}
    \Theta_c^{\text{within}},\\
    \mathcal{C}^{\text{cross}}
    &=
    \sum_{c=1}^{G}
    \frac{N_c}{N}
    \Theta_{c\leftarrow\bullet}^{\text{cross}}.
\end{align*}
Thus the system-wide measures weight cluster-level shares by the number of nodes in each receiving cluster, whereas the cluster-level tables report each cluster on its own row-normalized scale. In the empirical application, Tables~\ref{tab:clustered}--\ref{tab:clustered_ctrl_sep} report the cluster-level measures, while Table~\ref{tab:agg_conn} and Figure~\ref{fig:roll_idx} report the system-wide measures.

The quantity $Net_c$ plays a central role in the analysis that follows. Because net connectedness summarizes the overall transmission role of a cluster within the network, it provides a natural object through which to study the effect of identification on connectedness measurement. Later in this section, we use the distribution of $Net_c$ across admissible cluster orderings to measure ordering sensitivity within the BDY framework.

\subsection{Identification Robustness}
	\label{subsec:ordering}

	The BDY framework substantially reduces the identification burden associated with orthogonalized connectedness analysis. By grouping individual nodes into economically meaningful clusters, the number of admissible identification strategies falls from $N!$ to $C!$, often transforming an infeasible problem into a computationally tractable one. The reduction, however, is not complete elimination. Connectedness measures continue to depend on the ordering of clusters, and the resulting ordering sensitivity may affect both economic interpretation and any structural or causal interpretation based on the chosen identification.
	
	The central issue is that connectedness measures may attribute to a cluster variation that is not uniquely attributable to shocks originating from that cluster. Because orthogonalization allocates contemporaneous correlation according to an ordering, a cluster appearing early in the identification sequence may absorb variation that is common to several clusters. Measured connectedness can therefore reflect not only underlying shock transmission but also the particular identification strategy used to separate contemporaneous innovations.
	
	These observations highlight that the remaining identification problem is fundamentally one of robustness rather than computation. Clustering makes orthogonalized connectedness analysis feasible, but it does not guarantee that the resulting measures are insensitive to economically reasonable identification choices. This distinction motivates the diagnostic developed in the next subsection.
	
	To understand the source of this ordering sensitivity, consider the decomposition
	\begin{align*}
		u_t = \Lambda z_t + \eta_t, 
	\end{align*}
	where $z_t$ denotes common shocks, $\eta_t$ denotes idiosyncratic innovations originating within the network, and
	$\operatorname{Cov}(z_t,\eta_t)=0$. Examples of $z_t$ include movements in global risk appetite, volatility conditions, or interest-rate expectations that simultaneously affect multiple clusters. The covariance matrix of reduced-form shocks can then be written as
	\begin{align*}
		\Sigma = \Lambda \Sigma_z \Lambda' + \Sigma_\eta. 
	\end{align*}
	The first component captures common variation shared across clusters, while the second captures innovations internal to the network itself.
	
	This decomposition illustrates the source of ordering sensitivity. When common shocks are present, the orthogonalization procedure must attribute their contemporaneous variation to one or more clusters. Which cluster receives that variation depends on the ordering. A cluster positioned early in the orthogonalization sequence can absorb a larger share of the common component and therefore appear more important as a transmitter of shocks. Conversely, clusters positioned later can inherit less of that variation and appear less central. Ordering sensitivity therefore arises not because the underlying network changes, but because common variation is allocated differently across admissible identification strategies.
	
	The remainder of this section develops a diagnostic for measuring this ordering sensitivity and proposes an extension of the BDY framework that substantially mitigates it.
	
	\subsection{An Ordering-Sensitivity Diagnostic}
	\label{subsec:diagnostic}
	
	The existence of residual ordering sensitivity raises an important practical question: how large is the problem in applications? To answer that question, we introduce a diagnostic that quantifies the sensitivity of connectedness measures across admissible cluster orderings. This diagnostic also provides the benchmark against which the effectiveness of the proposed control-variable extension will later be evaluated.
	
	The discussion above suggests a simple implication. If economically relevant conclusions about a cluster's transmission role are robust to admissible orderings, its measured connectedness should not vary materially with its position in the orthogonalization ordering. Conversely, if connectedness varies systematically with ordering position, then part of the measured connectedness may reflect the attribution of common variation induced by the recursive identification rather than only underlying network transmission.
	
	To quantify this effect, fix a cluster $c$ and let $\Pi_{-c}$ denote the set of all $(C-1)!$ permutations of the remaining clusters. For any $\pi \in \Pi_{-c}$ and any position $p \in \{1,\ldots,C\}$, let $\mathit{Net}_{c,\pi}^{(p)}$ denote the net connectedness of cluster $c$ when $c$ is inserted in position $p$ while the relative ordering of all other clusters is held fixed at $\pi$. Thus, for each position $p$, the collection
	\begin{equation*}
		\left\{
		\mathit{Net}_{c,\pi}^{(p)}:
		\pi \in \Pi_{-c}
		\right\}
	\end{equation*}
	contains $(C-1)!$ net-connectedness values.
	
	For any two positions $p$ and $q$, we pair the corresponding orderings by holding $\pi$ fixed and define
	\begin{equation}
		\Delta_{c,\pi}^{(p,q)}=
		\mathit{Net}_{c,\pi}^{(p)}-
		\mathit{Net}_{c,\pi}^{(q)} .
		\label{eq:Delta_pq}
	\end{equation}
	The quantity $\Delta_{c,\pi}^{(p,q)}$ therefore measures the change in the estimated net connectedness of cluster $c$ when it moves from position $q$ to position $p$, holding fixed the relative ordering of every other cluster. The collection
	\begin{equation*}
		\left\{
		\Delta_{c,\pi}^{(p,q)}:
		\pi \in \Pi_{-c}
		\right\}
	\end{equation*}
	provides the distribution of these paired ordering effects.
	
	If connectedness were completely invariant to ordering, then for all positions $p$ and $q$,
	\begin{equation*}
		\mathbb{E}_{\pi}\!\left[\Delta_{c,\pi}^{(p,q)}\right] \approx 0,
		\qquad
		\mathbb{V}_{\pi}\!\left[\Delta_{c,\pi}^{(p,q)}\right] \approx 0,
	\end{equation*}
	where the expectation and variance are taken with respect to the uniform distribution over $\pi \in \Pi_{-c}$. Departures from these benchmarks reveal two distinct forms of identification sensitivity. A non-zero mean indicates a systematic ordering effect, whereas a large variance indicates heterogeneity in that effect across the admissible relative orderings of the remaining clusters.
	
	We focus on the especially informative first-versus-last comparison. For each $\pi \in \Pi_{-c}$, define
	\begin{equation}
		\Delta_{c,\pi}^{(1,C)}=
		\mathit{Net}_{c,\pi}^{(1)}-
		\mathit{Net}_{c,\pi}^{(C)} .
		\label{eq:first_last}
	\end{equation}
	\noindent This statistic compares the two extreme positions while holding fixed the relative ordering of all other clusters. Its distribution across $\pi$ therefore provides a natural robustness diagnostic: the farther the distribution lies from zero, and the more dispersed it is, the more sensitive the measured net connectedness of cluster $c$ is to recursive ordering. For scalar summaries, we use the mean paired difference
	\begin{equation*}
		\overline{\Delta}_{c}^{(1,C)}
		=
		\mathbb{E}_{\pi}\!\left[\Delta_{c,\pi}^{(1,C)}\right].
	\end{equation*}

	\subsection{Control Variables}
	\label{subsec:controls}
	
	The ordering-sensitivity diagnostic establishes both the existence and magnitude of the remaining identification problem. We now introduce the paper's control-variable extension, whose purpose is to reduce that sensitivity while preserving the computing benefits of the clustered framework.

	The discussion above suggests that ordering sensitivity can arise because common shocks are attributed to individual clusters during orthogonalization. A natural approach is therefore to condition on observable proxies for those common shocks before measuring connectedness among the bank clusters.
	Suppose that observable proxies for the common factors $z_t$ are available. Examples include measures of global and regional financial conditions, such as implied-volatility indices, interest-rate factors, or other indicators of market-wide risk. 
	
	The role of the control variables is not to model additional connectedness within the banking network. Instead, they are intended to proxy for common macro-financial shocks whose associated variation is removed before connectedness is measured among the bank clusters. By construction, the resulting connectedness measures are less influenced by shared exposure to the observed macro-financial controls. Under the maintained recursive identification in which innovations in the control cluster are contemporaneously exogenous to innovations in the bank clusters, the remaining cross-cluster connectedness can be interpreted as bank-to-bank transmission net of those observed common-factor shocks.
	
	Two approaches are possible. The first places control variables within the existing regional clusters. This specification absorbs some common variation, but it does not distinguish between shocks associated with the controls and shocks originating from the banks themselves. Consequently, part of the control-related common variation can continue to enter measured bank connectedness and may still affect ordering sensitivity. Additionally, correlation among the control variables can itself contribute to measured cross-cluster connectedness.
	
	The second approach, which we adopt, treats the controls as a separate cluster. This maintains an explicit distinction between innovations in the controls and innovations in the bank clusters, allowing connectedness to be interpreted conditional on the observed common factors. Under the maintained recursive exogeneity assumption described above, it also supports a causal interpretation of the remaining cross-cluster bank transmission. As we show below, this specification substantially reduces ordering sensitivity relative to the alternative.
	
	A potential concern is that creating an additional cluster might, at first sight, increase the number of possible identification strategies. To keep the number of identification strategies unchanged while retaining the possibility of causal interpretation under the maintained recursive assumptions, let $C_B$ denote the number of bank clusters. The augmented system then contains $C_B+1$ clusters: the control variables form cluster 1, and the bank clusters are indexed by $2,\ldots,C_B+1$. Partition the reduced-form innovation vector and covariance matrix as
	\begin{equation*}
		\bm{u}_t =
		\begin{pmatrix}
			\bm{u}_{1t} \\
			\bm{u}_{Bt}
		\end{pmatrix},
		\qquad
		\bm{\Sigma} =
		\begin{pmatrix}
			\bm{\Sigma}_{11} & \bm{\Sigma}_{1B} \\
			\bm{\Sigma}_{B1} & \bm{\Sigma}_{BB}
		\end{pmatrix},
	\end{equation*}
	where $\bm{u}_{1t}$ contains the control-cluster innovations and $\bm{u}_{Bt}$ stacks the innovations in the $C_B$ bank clusters. Assuming $\bm{\Sigma}_{11}$ is nonsingular, define the bank innovations residualized with respect to the control cluster by
	\begin{equation}
		\bm{u}_{Bt}^{\perp}
		=
		\bm{u}_{Bt}
		-
		\bm{\Sigma}_{B1}\bm{\Sigma}_{11}^{-1}\bm{u}_{1t}.
		\label{eq:control_residuals}
	\end{equation}
	By construction,
	\begin{equation*}
		\operatorname{Cov}\!\left(\bm{u}_{Bt}^{\perp},\bm{u}_{1t}\right)=\bm{0},
	\end{equation*}
	and the covariance matrix of the residualized bank innovations is
	\begin{equation}
		\bm{\Sigma}_{BB\cdot 1}
		=
		\operatorname{Var}\!\left(\bm{u}_{Bt}^{\perp}\right)
		=
		\bm{\Sigma}_{BB}
		-
		\bm{\Sigma}_{B1}\bm{\Sigma}_{11}^{-1}\bm{\Sigma}_{1B}.
		\label{eq:schur_control}
	\end{equation}
	Thus $\bm{\Sigma}_{BB\cdot 1}$ is the Schur complement of $\bm{\Sigma}_{11}$ in $\bm{\Sigma}$. Equivalently, for any pair of bank clusters $k,l\in\{2,\ldots,C_B+1\}$, the corresponding residual covariance block is
	\begin{equation*}
		\bm{\Sigma}'_{kl}
		=
		\bm{\Sigma}_{kl}
		-
		\bm{\Sigma}_{k1}\bm{\Sigma}_{11}^{-1}\bm{\Sigma}_{1l}.
	\end{equation*}
	
	The identification then proceeds in two stages:
	\begin{enumerate}
		\item \textbf{Stage 1:} Residualize the bank innovations with respect to the control cluster using equation~\ref{eq:control_residuals}, yielding the control-residualized covariance matrix $\bm{\Sigma}_{BB\cdot 1}$ in equation~\ref{eq:schur_control}.
		\item \textbf{Stage 2:} Apply the usual BDY block orthogonalization to $\bm{\Sigma}_{BB\cdot 1}$ for each admissible ordering of the $C_B$ bank clusters.
	\end{enumerate}
	
	For a given ordering $\pi$ of the bank clusters, let $\bm{Q}_{B}(\pi)$ denote the corresponding BDY transformation, expressed in the original bank-variable order (and hence block lower-triangular when the bank clusters are arranged according to $\pi$), and let $\bm{\Omega}_{B}(\pi)$ denote the resulting block-diagonal covariance matrix, so that
	\begin{equation*}
		\bm{\Sigma}_{BB\cdot 1}
		=
		\bm{Q}_{B}(\pi)\bm{\Omega}_{B}(\pi)\bm{Q}_{B}(\pi)'.
	\end{equation*}
	The full control-first identification can then be written as
	\begin{equation}
		\bm{Q}_{C_B+1}(\pi)
		=
		\begin{pmatrix}
			\bm{I} & \bm{0} \\
			\bm{\Sigma}_{B1}\bm{\Sigma}_{11}^{-1} & \bm{Q}_{B}(\pi)
		\end{pmatrix},
		\qquad
		\bm{\Omega}_{C_B+1}(\pi)
		=
		\begin{pmatrix}
			\bm{\Sigma}_{11} & \bm{0} \\
			\bm{0} & \bm{\Omega}_{B}(\pi)
		\end{pmatrix},
		\label{eq:control_first_identification}
	\end{equation}
	which satisfies
	\begin{equation*}
		\bm{\Sigma}
		=
		\bm{Q}_{C_B+1}(\pi)\bm{\Omega}_{C_B+1}(\pi)\bm{Q}_{C_B+1}(\pi)'.
	\end{equation*}
	
	Crucially, adding a control-variable cluster adds one preliminary residualization step but \emph{does not increase the number of identification strategies}. The control cluster is fixed first, and only the $C_B$ bank clusters are permuted, so the number of orderings remains $C_B!$.
	
	We illustrate this for the case of $C_B = 3$ bank clusters, denoted by the vertical-, diagonal-, and horizontal-line boxes, with the control cluster denoted by the dotted box. Adding a fourth cluster in the usual way would raise the number of admissible orderings from $3! = 6$ to $4! = 24$. By contrast, fixing the control cluster in the first position and permuting only the three bank clusters preserves the six orderings. On the left we list the six permutations with the control cluster pinned first; on the right we show the equivalent system after the bank innovations have been residualized with respect to the control cluster, where the primed boxes denote residuals net of the control variables. In words, the inclusion of the control cluster contributes only a preliminary partialing out step before the bank clusters are placed in all $C_B!$ orderings.
	
	\begin{center}
		\begin{minipage}{0.34\textwidth}
			\begin{enumerate}[label={}]
				\item \dtbox{} \rule[-7pt]{0.5pt}{.5cm} \vlbox{} \dlbox{} \hlbox{}
				\item \dtbox{} \rule[-7pt]{0.5pt}{.5cm} \vlbox{} \hlbox{} \dlbox{}
				\item \dtbox{} \rule[-7pt]{0.5pt}{.5cm} \dlbox{} \hlbox{} \vlbox{}
				\item \dtbox{} \rule[-7pt]{0.5pt}{.5cm} \dlbox{} \vlbox{} \hlbox{}
				\item \dtbox{} \rule[-7pt]{0.5pt}{.5cm} \hlbox{} \dlbox{} \vlbox{}
				\item \dtbox{} \rule[-7pt]{0.5pt}{.5cm} \hlbox{} \vlbox{} \dlbox{}
			\end{enumerate}
		\end{minipage}
		\begin{minipage}{0.08\textwidth}
			\centering $\Rightarrow$
		\end{minipage}
		\begin{minipage}{0.34\textwidth}
			\begin{enumerate}[label={}]
				\item \vlbox{}$'$ \dlbox{}$'$ \hlbox{}$'$
				\item \vlbox{}$'$ \hlbox{}$'$ \dlbox{}$'$
				\item \dlbox{}$'$ \hlbox{}$'$ \vlbox{}$'$
				\item \dlbox{}$'$ \vlbox{}$'$ \hlbox{}$'$
				\item \hlbox{}$'$ \dlbox{}$'$ \vlbox{}$'$
				\item \hlbox{}$'$ \vlbox{}$'$ \dlbox{}$'$
			\end{enumerate}
		\end{minipage}
	\end{center}
	
	The role of the control cluster is therefore conceptual as well as computational. By removing variation associated with the observed controls first, the remaining identification problem is confined to shocks that are more plausibly specific to the clusters of substantive interest.
	
	As such we consider the same identification strategies as before, but conditional on the fact that components linearly associated with the control cluster have been removed. The resulting framework therefore preserves the tractability of clustered connectedness analysis while substantially reducing the remaining identification uncertainty due to ordering.
	
	\section{Global Banking}
	\label{sec:results}
	
	We now apply the framework to global banking. We first describe the data and empirical implementation, then characterize connectedness under alternative treatments of the controls, evaluate its sensitivity to recursive ordering, and finally synthesize the economic interpretation across identification strategies.
	
	\subsection{Data and Empirical Implementation}
	\label{sec:data}
	
	Our dataset comprises daily volatility measures for seventy-one banks grouped into seven regional clusters: (1) United States (fourteen banks), (2) Canada (six banks), (3) Europe (seventeen banks), (4) Europe Other (fourteen banks), (5) Australia (five banks), (6) Japan (eleven banks), and (7) China (four banks). The sample period runs from September 12, 2003 to February 20, 2024. We source daily open-high-low-close (OHLC) equity-price observations from Thomson Reuters and transform them into daily realized volatility measures following the range-based estimator of \citet{GarmanKlass1980}. Volatilities, rather than returns, are especially well suited to the study of banking connectedness: bank equity volatilities respond sharply to credit-risk and counterparty-risk news, and their cross-sectional co-movement is a well-established barometer of systemic stress \citep{Demirer2018}.
	  
	This dataset is well suited to our objectives for two reasons. First, it provides broad geographic coverage of the global banking system while preserving economically meaningful regional clusters. Second, the long daily sample allows us to estimate connectedness measures with sufficient precision to evaluate ordering sensitivity across alternative identification strategies.
	
	In our application we use four control variables for the full sample, each an implied-volatility index chosen to proxy for a distinct regional channel of common macro-financial volatility variation:
	\begin{itemize}
		\item \textbf{VIX:} the Chicago Board Options Exchange Volatility Index, a forward-looking implied-volatility measure on the S\&P 500. VIX proxies for U.S.\ market-wide risk.
		\item \textbf{VSTOXX:} the Euro STOXX 50 implied-volatility index, the European analog of the VIX, proxying for euro-area risk conditions.
		\item \textbf{VXJ:} the Nikkei 225 implied-volatility index, proxying for Japanese and broader Asian-Pacific risk conditions.
		\item \textbf{VHSI:} the Hang Seng Index implied-volatility index, proxying for Chinese and Hong Kong risk conditions.
	\end{itemize}
	Together, these four implied-volatility indices provide proxies for important regional dimensions of common macro-financial volatility variation: U.S., European, Japanese, and Chinese risk conditions. The VIX is obtained from FRED, VSTOXX from the STOXX Limited database, VXJ from the Center for the Study of Finance and Insurance at Osaka University, and VHSI from the Hong Kong Exchange.
	
	For the full sample we are unable to assign a dedicated control variable to the Canadian and Australian clusters, because the corresponding implied-volatility indices begin only in mid-2012. We therefore report a second specification over the subsample beginning July~9,~2012, for which two additional controls are available: the \textbf{TSX} implied-volatility index for Canada and the \textbf{ASX} implied-volatility index for Australia. In the subsample, each of the six bank clusters with an available control (U.S., Canada, Europe, Australia, Japan, and China) is matched to its own regional volatility index, so the control cluster includes proxies for all of the major regions in the network. We report the subsample aggregate results in Panel~B of Table~\ref{tab:agg_conn} below and use the full four-control set elsewhere.
	 
	We estimate a second-order VAR system equation-by-equation using an elastic net \citep{ZouHastie2005},
	\begin{equation}
		\hat{\bm{\beta}} = \underset{\bm{\beta}}{\operatorname{argmin}} \left( \sum_{t=1}^{T} \left( y_t - \sum_{i} \beta_i x_{it} \right)^2 + \lambda \sum_{i=1}^{K} \left( \frac{1}{2} |\beta_i| + \frac{1}{4} \beta_i^2 \right) \right),
		\label{eq:enet}
	\end{equation}
	where the tuning parameter $\lambda$ is selected equation-by-equation via ten-fold cross-validation. After estimating all equations, we collect the residuals and assemble in the usual way an estimate $\hat{\bm{\Sigma}}$ of the reduced-form shock covariance matrix.
	 
	With the VAR estimates in hand, we implement the identification procedures developed in Section~\ref{sec:framework}. As discussed in Subsection~\ref{subsec:diagnostic}, rather than commit to a single cluster ordering, we compute the variance decompositions associated with all $7!=5{,}040$ admissible cluster orderings.  We report both their average and the full distribution across orderings, the latter serving as our ordering-sensitivity diagnostic.  Throughout the empirical analysis, we adopt a variance decomposition horizon of $H=10$ days.
	
	\subsection{Global Banking Connectedness}
	\label{subsec:banking_connectedness}
	
	We begin by characterizing the connectedness structure itself, moving from the baseline clustered specification to alternative treatments of the controls, aggregate comparisons, network visualization, and time variation.
	
	\subsubsection{Clustered Connectedness Without Controls}
	
	We begin with the baseline clustered specification without control variables. This benchmark establishes the connectedness structure that emerges when common macro-financial variation is not separately accounted for. The subsequent specifications will show how that picture changes once observable common factors are introduced explicitly.

	The baseline clustered specification reveals a clear hierarchy of volatility transmission across the global banking system. The United States is the dominant net transmitter of volatility shocks, with a net connectedness of $+59.48$ percentage points ($\text{To}=82.24$, $\text{From}=22.76$), and Europe is a modest net transmitter ($+8.01$). Every other cluster is a net receiver: Australia is the largest ($-33.16$), followed by Canada ($-22.92$), while Europe Other ($-4.09$), China ($-4.17$), and Japan ($-3.15$) receive comparatively little on net. Canada's position reflects its tight linkage to the U.S.\ banking system, with $24.30\%$ of Canadian forecast error variance attributable to U.S.\ banks. China and Japan exhibit very low overall connectedness; their own-cluster shares of $92.23\%$ and $88.22\%$ indicate that their banking systems are comparatively insulated from the rest of the global network. Table~\ref{tab:clustered} reports the corresponding cluster-level FEVD matrix.

	Within the European clusters, the strong mutual spillovers between EUR and EUO ($24.71\%$ from EUR to EUO; $22.13\%$ from EUO to EUR) are the strongest pairwise cluster linkages in the table and are consistent with tight financial integration.
	
\begin{table}[tb!]
\centering
\caption{Clustered Connectedness: Without controls}
\label{tab:clustered}
\smallskip
%		\small
\setlength{\tabcolsep}{6pt}   % default is 6pt; shrink inter-column space, not the font
\begin{tabular}{llrrrrrrr|r}
    \toprule
    &&\multicolumn{8}{c}{From}\\
    &&USA &CAN &EUR &EUO &AUS &JPN &CHN & Total \\ \midrule
    \multirow{8}{*}{\rotatebox{90}{To}}
    &USA&77.24 & 4.90 & 8.43 & 7.09 & 1.57 & 0.60 & 0.16 & 22.76 \\
    &CAN&24.30 & 58.65 & 6.40 & 6.20 & 2.34 & 1.32 & 0.80 & 41.35 \\
    &EUR&15.89 & 2.30 & 57.10 & 22.13 & 1.33 & 0.82 & 0.42 & 42.90 \\
    &EUO&18.32 & 3.10 & 24.71 & 51.45 & 1.28 & 0.79 & 0.34 & 48.55 \\
    &AUS&17.94 & 5.62 & 7.84 & 6.67 & 58.47 & 2.83 & 0.62 & 41.53 \\
    &JPN&4.56 & 1.27 & 1.99 & 1.54 & 1.16 & 88.22 & 1.26 & 11.78 \\
    &CHN&1.22 & 1.23 & 1.54 & 0.84 & 0.67 & 2.26 & 92.23 & 7.77 \\ \midrule
    &Total &82.24 & 18.43 & 50.91 & 44.46 & 8.37 & 8.63 & 3.60 &  \\ \midrule
    &Net   &59.48 & -22.92 & 8.01 & -4.09 & -33.16 & -3.15 & -4.17 &  \\
    \bottomrule
\end{tabular}
\end{table}

	Having established the baseline connectedness structure, we now examine how that structure changes when common regional risk factors are accounted for through the control-variable framework developed in Subsection~\ref{subsec:controls}.
	
	\subsubsection{Clustered Connectedness With Controls}
	
	Introducing control variables substantially changes the measured pattern of connectedness by separating variation associated with observed regional risk factors from bank-cluster spillovers. When the controls are distributed within their corresponding regional clusters, the resulting cluster-level FEVD matrix (Table~\ref{tab:clustered_ctrl}) retains the seven-cluster structure, but the control-related variation is absorbed within the regional clusters, increasing own-cluster shares and correspondingly reducing measured cross-cluster spillovers. Relative to the no-controls specification, the own-cluster share of every cluster except China rises (for example, the U.S.\ from $77.24\%$ to $83.97\%$ and EUR from $57.10\%$ to $65.03\%$). China is the exception: its own-cluster share falls from $92.23\%$ to $84.21\%$, reflecting the additional linkages that its control, the VHSI, brings into the Chinese cluster. The largest net positions compress (the U.S.\ from $+59.48$ to $+46.01$, Canada from $-22.92$ to $-15.96$, Australia from $-33.16$ to $-21.33$), while the European clusters and China move modestly in the opposite direction (EUR from $+8.01$ to $+11.75$, EUO from $-4.09$ to $-8.34$, China from $-4.17$ to $-9.08$). The ranking is unchanged: the U.S.\ remains the dominant net transmitter, Europe remains a modest net transmitter, and the remaining clusters remain net receivers, led by Australia and Canada.
	
\begin{table}[tb!]
\centering
\caption{Clustered Connectedness: Controls in clusters}
\label{tab:clustered_ctrl}
\smallskip
\setlength{\tabcolsep}{6pt}   
\begin{tabular}{llrrrrrrr|r}
    \toprule
    &&\multicolumn{8}{c}{From}\\
    &&USA &CAN &EUR &EUO &AUS &JPN &CHN & Total \\ \midrule
    \multirow{8}{*}{\rotatebox{90}{To}}
    &USA&83.97 & 3.28 & 6.98 & 3.95 & 0.62 & 0.58 & 0.63 & 16.03 \\
    &CAN&16.82 & 72.02 & 4.46 & 2.85 & 1.20 & 1.03 & 1.63 & 27.98 \\
    &EUR&12.34 & 1.31 & 65.03 & 18.91 & 0.71 & 0.89 & 0.81 & 34.97 \\
    &EUO&12.45 & 1.53 & 24.20 & 59.58 & 0.58 & 0.80 & 0.86 & 40.42 \\
    &AUS&10.28 & 3.38 & 4.91 & 3.38 & 74.17 & 2.69 & 1.19 & 25.83 \\
    &JPN&4.36 & 0.86 & 2.55 & 1.30 & 0.76 & 88.57 & 1.60 & 11.43 \\
    &CHN&5.78 & 1.66 & 3.62 & 1.68 & 0.64 & 2.41 & 84.21 & 15.79 \\ \midrule
    &Total &62.04 & 12.02 & 46.72 & 32.08 & 4.50 & 8.39 & 6.71 &  \\ \midrule
    &Net   &46.01 & -15.96 & 11.75 & -8.34 & -21.33 & -3.04 & -9.08 &  \\
    \bottomrule
\end{tabular}
\end{table}

\begin{table}[tb!]
\centering
\caption{Clustered Connectedness: Separate control cluster}
\label{tab:clustered_ctrl_sep}
\smallskip
\setlength{\tabcolsep}{6pt}   
\begin{tabular}{llrrrrrrrr|r}
    \toprule
    &&\multicolumn{9}{c}{From}\\
    &&CTL &USA &CAN &EUR &EUO &AUS &JPN &CHN & Total \\ \midrule
    \multirow{9}{*}{\rotatebox{90}{To}}
    &CTL&95.66 & 1.43 & 0.24 & 0.81 & 0.78 & 0.18 & 0.54 & 0.36 & 4.34 \\
    &USA&9.27 & 80.42 & 2.29 & 4.08 & 2.76 & 0.55 & 0.43 & 0.19 & 19.58 \\
    &CAN&11.16 & 10.07 & 72.36 & 1.88 & 1.68 & 1.03 & 0.69 & 1.11 & 27.64 \\
    &EUR&9.32 & 6.84 & 0.72 & 64.41 & 16.97 & 0.61 & 0.58 & 0.55 & 35.59 \\
    &EUO&10.11 & 7.07 & 0.85 & 20.29 & 60.07 & 0.47 & 0.58 & 0.55 & 39.93 \\
    &AUS&7.93 & 5.60 & 2.41 & 2.77 & 2.28 & 76.30 & 1.99 & 0.73 & 23.70 \\
    &JPN&4.04 & 2.05 & 0.43 & 1.04 & 0.71 & 0.64 & 90.20 & 0.88 & 9.80 \\
    &CHN&1.36 & 0.70 & 1.11 & 1.80 & 0.92 & 0.55 & 1.21 & 92.35 & 7.65 \\ \midrule
    &Total &53.19 & 33.76 & 8.05 & 32.68 & 26.10 & 4.03 & 6.04 & 4.38 &  \\ \midrule
    &Net   &48.85 & 14.19 & -19.59 & -2.90 & -13.83 & -19.68 & -3.76 & -3.28 &  \\
    \bottomrule
\end{tabular}
\end{table}
	
	Gathering the control variables into a separate cluster changes the measured roles throughout the network. Table~\ref{tab:clustered_ctrl_sep} shows that the control cluster is by far the largest net transmitter ($+48.85$), reflecting the fact that control-related common variation is attributed to that cluster under the control-first identification, while it receives almost nothing in return ($\text{From}=4.34$). The U.S.\ net position falls from $+59.48$ without controls to $+14.19$, a decline of roughly three quarters. Under the maintained control-first identification, this large reduction indicates that a substantial portion of the transmission attributed to the U.S.\ in the no-controls specification is associated with variation captured by the observed common macro-financial controls. Yet the U.S.\ remains the largest net transmitter among the bank clusters, and indeed the only bank cluster with positive net connectedness in this specification. Europe moves to near balance ($-2.90$), and the remaining clusters are net receivers of moderate size: Australia ($-19.68$), Canada ($-19.59$), Europe Other ($-13.83$), Japan ($-3.76$), and China ($-3.28$). China is the cluster least exposed to the controls, with only $1.36\%$ of its forecast error variance attributable to the control cluster, consistent with its comparatively limited measured exposure to the observed global macro-financial factors.
	
	The preceding analysis focused on cluster-level connectedness under alternative treatments of the control variables. We now step back and compare the system-wide implications of the clustered and generalized identification frameworks, using the distinction developed in Subsection~\ref{subsec:connectedness_measures}.

	\subsubsection{Aggregate Connectedness: Generalized vs. Clustered}
	
	The system-wide connectedness measures reinforce a central conclusion of the paper: treating observed common factors more explicitly primarily reduces the cross-group component of measured connectedness while leaving the within-group component largely unchanged. Throughout this subsection, total, within-cluster, and cross-cluster connectedness refer to the system-wide averages defined in Subsection~\ref{subsec:connectedness_measures}, rather than to the cluster-level To, From, and Net measures in Tables~\ref{tab:clustered}--\ref{tab:clustered_ctrl_sep}. Table~\ref{tab:agg_conn} summarizes these system-wide measures under three specifications: the generalized approach with controls, the clustered approach with controls placed within the existing clusters, and the clustered approach with controls gathered into a separate control cluster. For the generalized specification, the within/cross classification is a descriptive reporting decomposition of the generalized FEVD; generalized identification itself still contains a single shock block.

	\begin{table}[tb!]
		\centering
		\caption{System-Wide Aggregate Connectedness Across Specifications}
		\label{tab:agg_conn}
		\smallskip
		
		\textbf{Panel A: Full sample (2003--2024), four controls}
		\smallskip
		
		\begin{tabular}{lccc}
			\toprule
			& Generalized approach & Clustered approach  & Clustered approach  \\
			&(with controls) & (ctrls. in clu.) & (sep. ctrl. clu.)\\
			\midrule
			Total con. &  76.49 & 67.34& 65.38 \\
			Within&  39.36 & 41.35 & 40.34  \\
			Cross&  37.13& 25.99 & 25.04 \\
			\bottomrule
		\end{tabular}
		
		\bigskip
		
		\textbf{Panel B: Subsample (July 9, 2012--2024), six controls}
		\smallskip
		
		\begin{tabular}{lccc}
			\toprule
			& Generalized approach & Clustered approach  & Clustered approach  \\
			&(with controls) & (ctrls. in clu.) & (sep. ctrl. clu.)\\
			\midrule
			Total  con. &76.20  & 66.81& 65.88 \\
			Within&  41.57 & 43.10 & 43.21  \\
			Cross&  34.63& 23.71 & 22.67 \\
			\bottomrule
		\end{tabular}
		
		\smallskip
		{\footnotesize \emph{Notes:} Panel~A uses the four full-sample controls (VIX, VSTOXX, VXJ, VHSI). Panel~B uses the subsample beginning July~9,~2012, over which two additional controls become available (the TSX implied-volatility index for Canada and the ASX implied-volatility index for Australia) so that each of the six regions USA, Canada, Europe, Australia, Japan, and China is matched to its own regional volatility index.}
	\end{table}

	Three findings stand out. First, total connectedness falls as the controls are given an increasingly explicit role in the identification: it is $76.49$ under the generalized approach with controls, $67.34$ when the controls are placed within the existing clusters, and $65.38$ when they are gathered into a separate cluster. Second, within-cluster connectedness is very stable across the three specifications, moving only between $39.36$ and $41.35$, because each specification preserves within-cluster correlations. Third, the cross-cluster component carries essentially all of the difference: it drops from $37.13$ under the generalized approach to $25.99$ with controls in clusters and $25.04$ with a separate control cluster. This pattern is consistent with the control cluster's intended role: variation associated with the observed common factors that would otherwise contribute to measured cross-cluster connectedness is instead attributed to the controls.
	
	A further implication is that moving the controls from inside the clusters to a separate cluster lowers total connectedness only marginally (from $67.34$ to $65.38$) but is what delivers the large reduction in ordering sensitivity documented below. The aggregate levels are therefore similar across the two control specifications, while their identification properties differ sharply.
	
	Panel~B of Table~\ref{tab:agg_conn} repeats the exercise over the July~2012--2024 subsample, for which Canadian (TSX) and Australian (ASX) implied-volatility controls are also available, so that every major region is matched to its own control. The same ordering across specifications holds. 
When the additional controls are placed within the existing clusters, total connectedness ($66.81$) and cross-cluster connectedness ($23.71$) lie only slightly above the separate-cluster values ($65.88$ and $22.67$), and both are well below the generalized approach ($76.20$ and $34.63$). As in the full sample, the two placements deliver similar aggregate levels. When instead the full set of controls is gathered into a separate cluster, cross-cluster connectedness falls to $22.67$ (well below the within-cluster value of $43.21$), leaving a substantially smaller across-cluster component after conditioning on the observed controls. These results reinforce the central message of the paper: a richer separate control cluster can account for more observed common variation, thereby reducing measured cross-cluster connectedness and substantially improving identification robustness. Taken together, these findings show that the proposed control-variable framework modifies connectedness primarily where identification uncertainty is greatest, i.e. the cross-cluster component, while leaving the economically meaningful within-cluster structure largely intact.

	\begin{figure}[p!]
		\centering
		\includegraphics[width=0.68\textwidth]{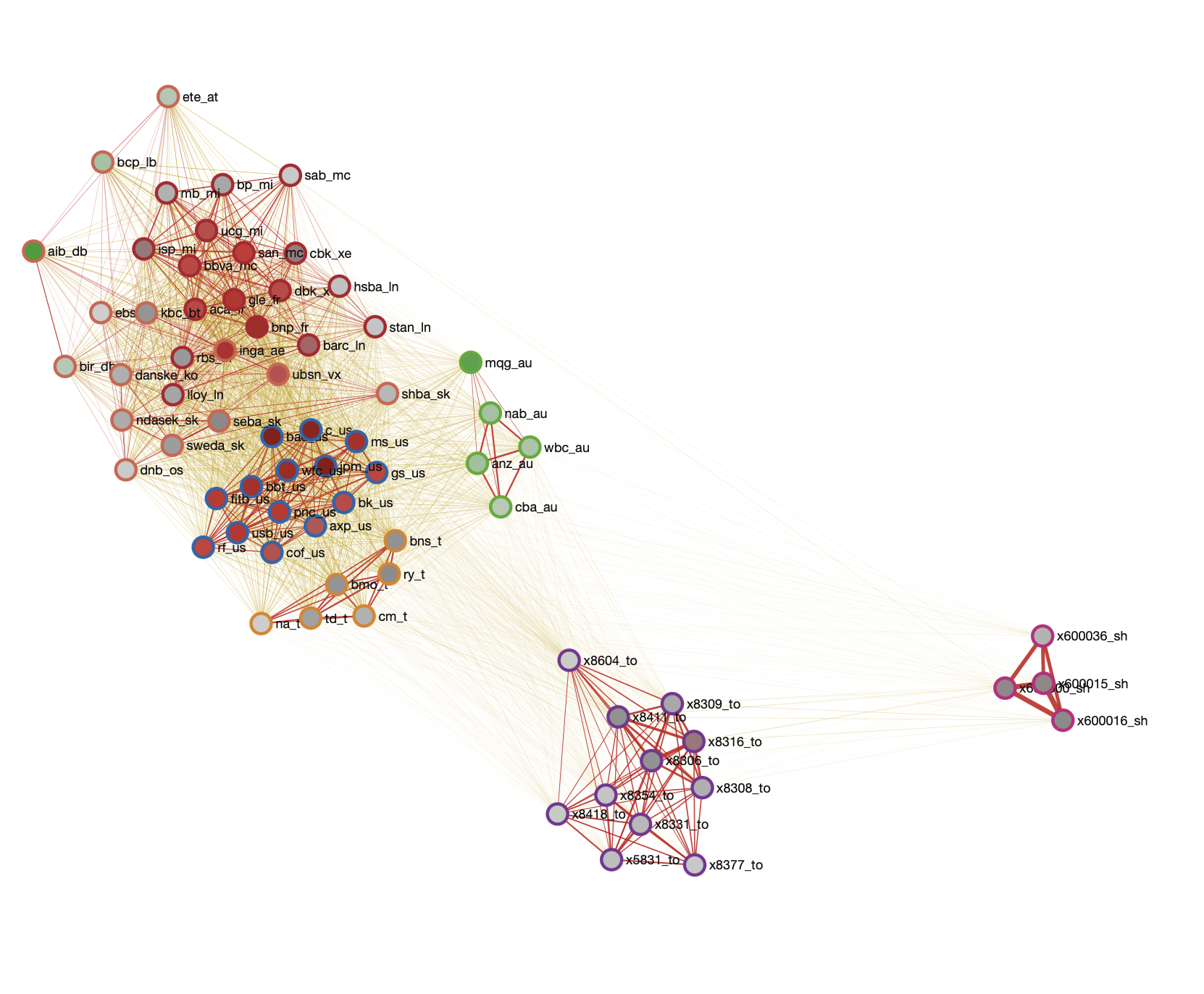}
		\caption{Network graph: Generalized identification (without controls).}
		\label{fig:network_gen_noctrl}
		\vspace{-0.2em}
		\includegraphics[width=0.68\textwidth]{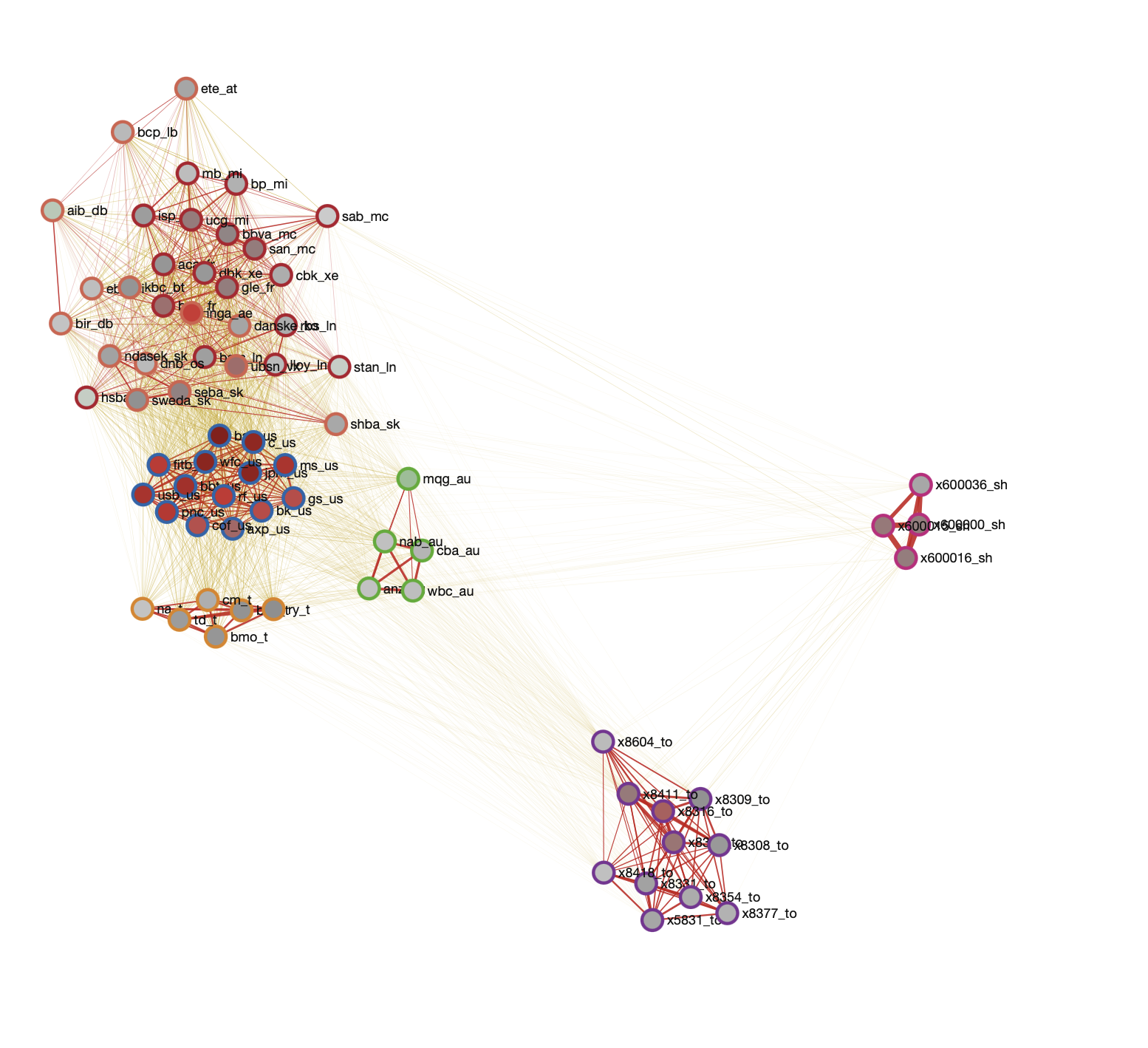}
		\caption{Network graph: Clustered identification (without controls).}
		\label{fig:network_clu_noctrl}
	\end{figure}

	\begin{figure}[p!]
		\centering
		\includegraphics[width=0.68\textwidth]{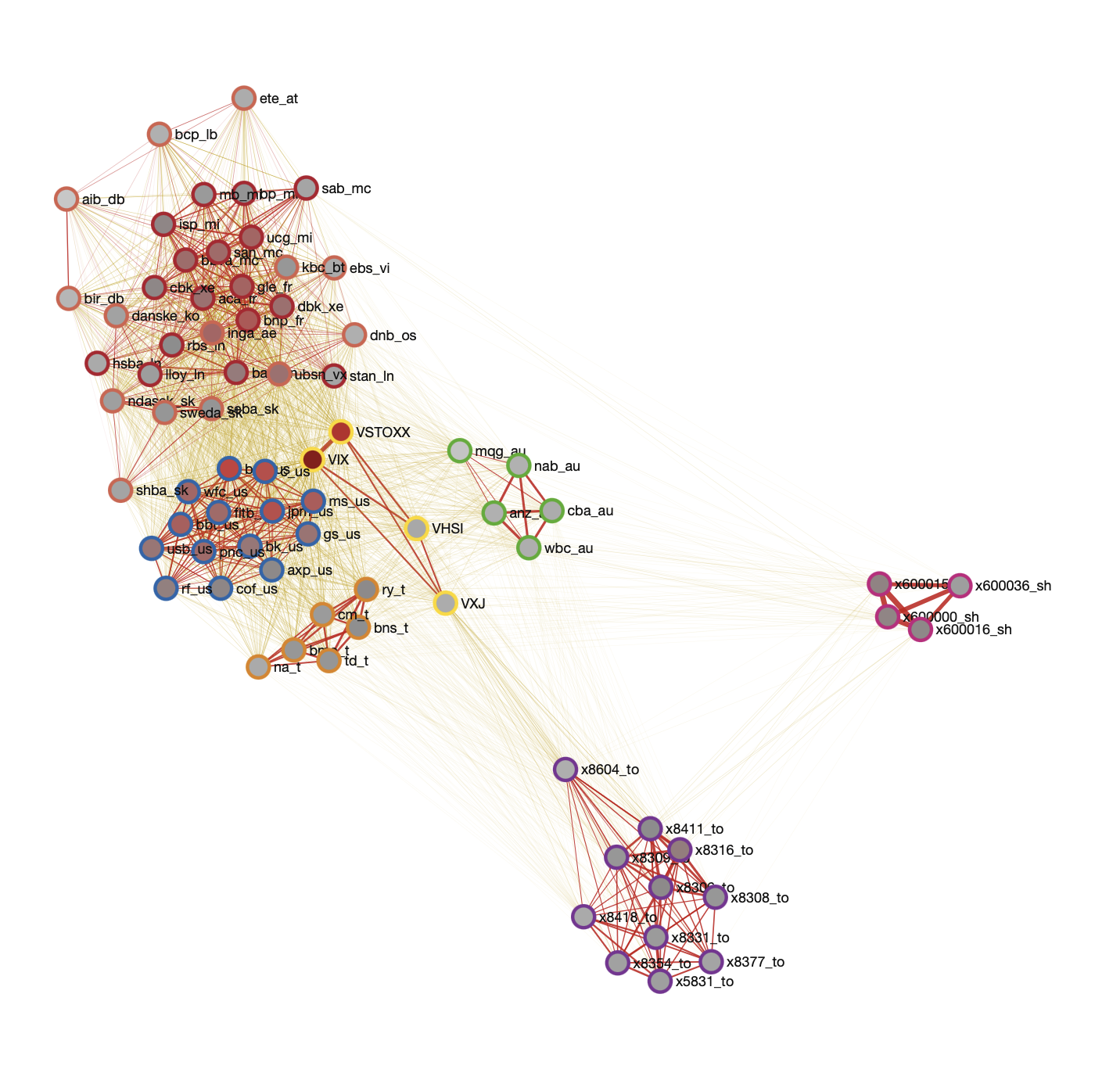}
		\caption{Network graph: Generalized identification (with four controls: VIX, VSTOXX, VXJ, VHSI).}
		\label{fig:network_gen_ctrl}
		\vspace{-0.5em}
		\includegraphics[width=0.68\textwidth]{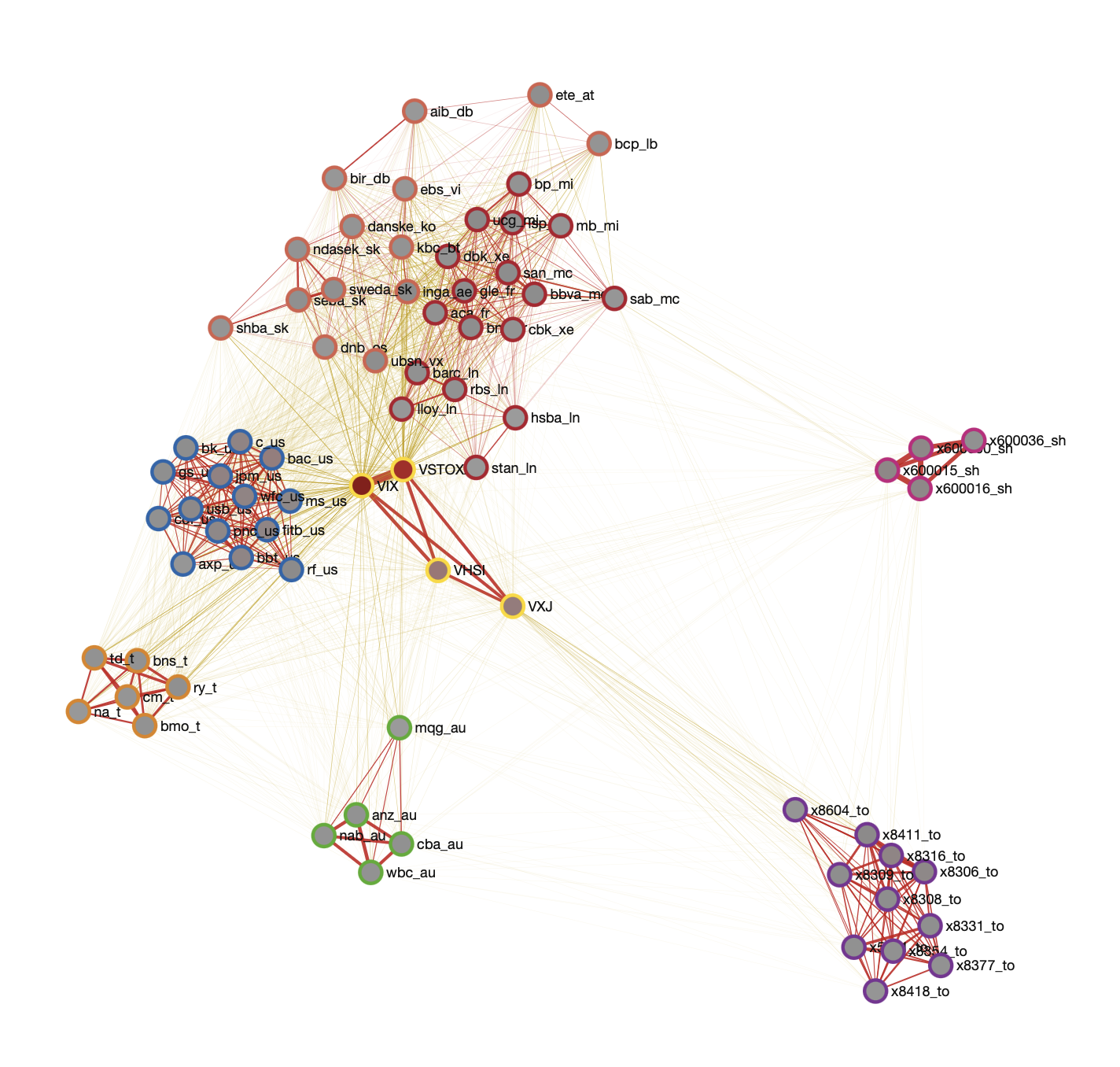}
		\caption{Network graph: Clustered identification (with four controls: VIX, VSTOXX, VXJ, VHSI).}
		\label{fig:network_clu_ctrl}
	\end{figure}

	\begin{figure}[t!]
		\centering
		\includegraphics[width=0.68\textwidth]{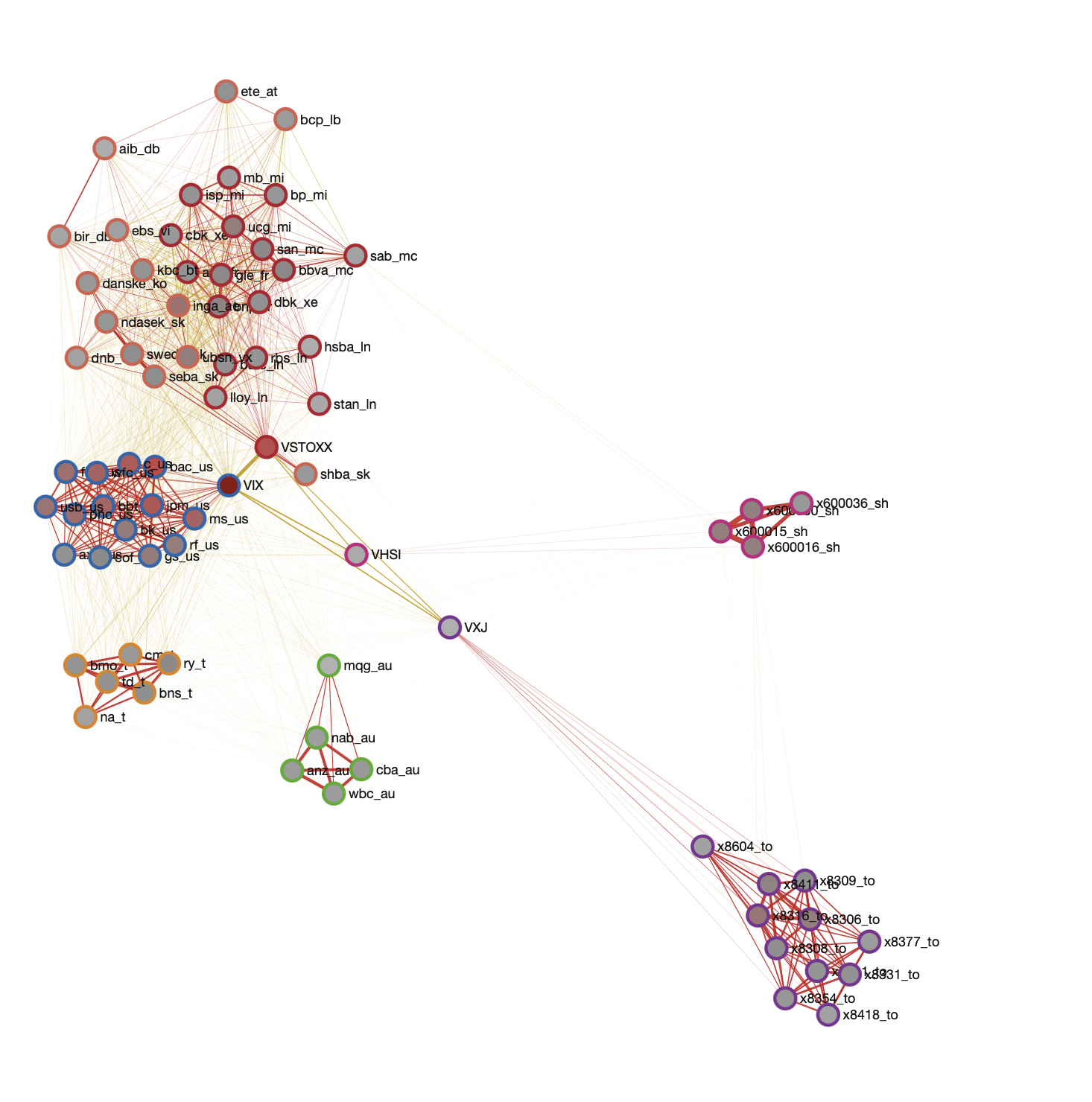}
		\caption{Network graph: Clustered identification (four controls ordered with the corresponding region: VIX, VSTOXX, VXJ, VHSI).}
		\label{fig:network_clu_ctrl_reg}
	\end{figure}

	\subsubsection{Network Visualization}
	
	The connectedness tables quantify the strength and direction of spillovers. Network visualizations complement those tables by revealing the overall topology of the banking system and illustrating how alternative identification strategies reshape the network.
	
	Figures~\ref{fig:network_gen_noctrl}--\ref{fig:network_clu_ctrl_reg} compare the generalized and clustered identification strategies, both without controls and under alternative treatments of the control variables. Node colors reflect bank-level net connectedness (red shades indicate net transmitters; green shades indicate net receivers), with darker shades indicating larger magnitudes. Node borders identify the economic reporting groups. Edge colors classify links as within-group (red) or cross-group (yellow); under clustered identification this coincides with within-cluster co-movement versus cross-cluster connectedness, whereas under generalized identification the classification is descriptive because contemporaneous shocks are not orthogonalized across the reporting groups. All network layouts are constructed using the ForceAtlas2 algorithm \citep{Jacomy2014} with equal node sizes.
	
	The generalized identification produces a densely connected banking network in which cross-group linkages dominate the visual topology. Figure~\ref{fig:network_gen_noctrl} shows that yellow (cross-group) edges dominate because the generalized approach retains contemporaneous correlation across nodes, producing comparatively dense pairwise linkages. European banks (EUR and EUO clusters, red borders) occupy the center of the network together with U.S.\ banks (blue borders), whereas Japan and China remain clearly separated on the periphery.

	The clustered identification produces a visibly different network topology by shifting connectedness from across-cluster to within-cluster relationships. Figure~\ref{fig:network_clu_noctrl} shows that within-cluster red edges become more prominent while across-cluster yellow edges thin out, reflecting the reassignment of within-cluster co-movement to the diagonal FEVD blocks. The U.S.\ cluster remains centrally located, consistent with its role as the dominant net transmitter in the baseline clustered specification.

	Introducing the four control variables substantially changes the role played by observed common macro-financial factors in the network. Figures~\ref{fig:network_gen_ctrl}--\ref{fig:network_clu_ctrl_reg} illustrate how the resulting topology depends on the treatment of those controls.  In the first two of the three figures, VIX and VSTOXX occupy central positions with thick outgoing linkages to nearly every bank cluster, consistent with their role as important common-factor proxies, while VXJ and VHSI play more localized roles, connecting primarily to Japanese and Chinese banks respectively.
	
	The three treatments differ sharply in how much the controls reshape the network. Under the generalized identification (Figure~\ref{fig:network_gen_ctrl}), adding the controls as a separate cluster leaves the relative positions of the seven bank clusters essentially unchanged: because the generalized approach does not orthogonalize contemporaneous shocks, the controls are layered onto the existing structure without being assigned a recursive priority. Under the clustered identification with the controls in their own cluster (Figure~\ref{fig:network_clu_ctrl}), by contrast, the control variables are drawn to the center of the network and substantially rearrange the relative positions of all seven bank clusters, reflecting the control-first construction: the bank innovations are first residualized with respect to the control cluster, so variation linearly associated with the observed controls is attributed to that cluster before the bank clusters are orthogonalized. Finally, when the four regional controls are instead placed at the top of their \emph{corresponding} regional clusters (Figure~\ref{fig:network_clu_ctrl_reg}), they cease to act as system-wide transmitters: each control's associated common-factor variation is absorbed within its own regional cluster rather than attributed to a separate system-wide control cluster, so the controls recede from the center and generate little connectedness to the rest of the network.

	The previous results are based on the full sample. We now examine how connectedness evolves over time to determine whether the patterns identified above are stable or concentrated in particular episodes.

	\subsubsection{Rolling-Window Analysis}
	
\begin{figure}[H]
		\centering
		\begin{subfigure}{0.70\textwidth}
			\centering
			\includegraphics[width=\linewidth]{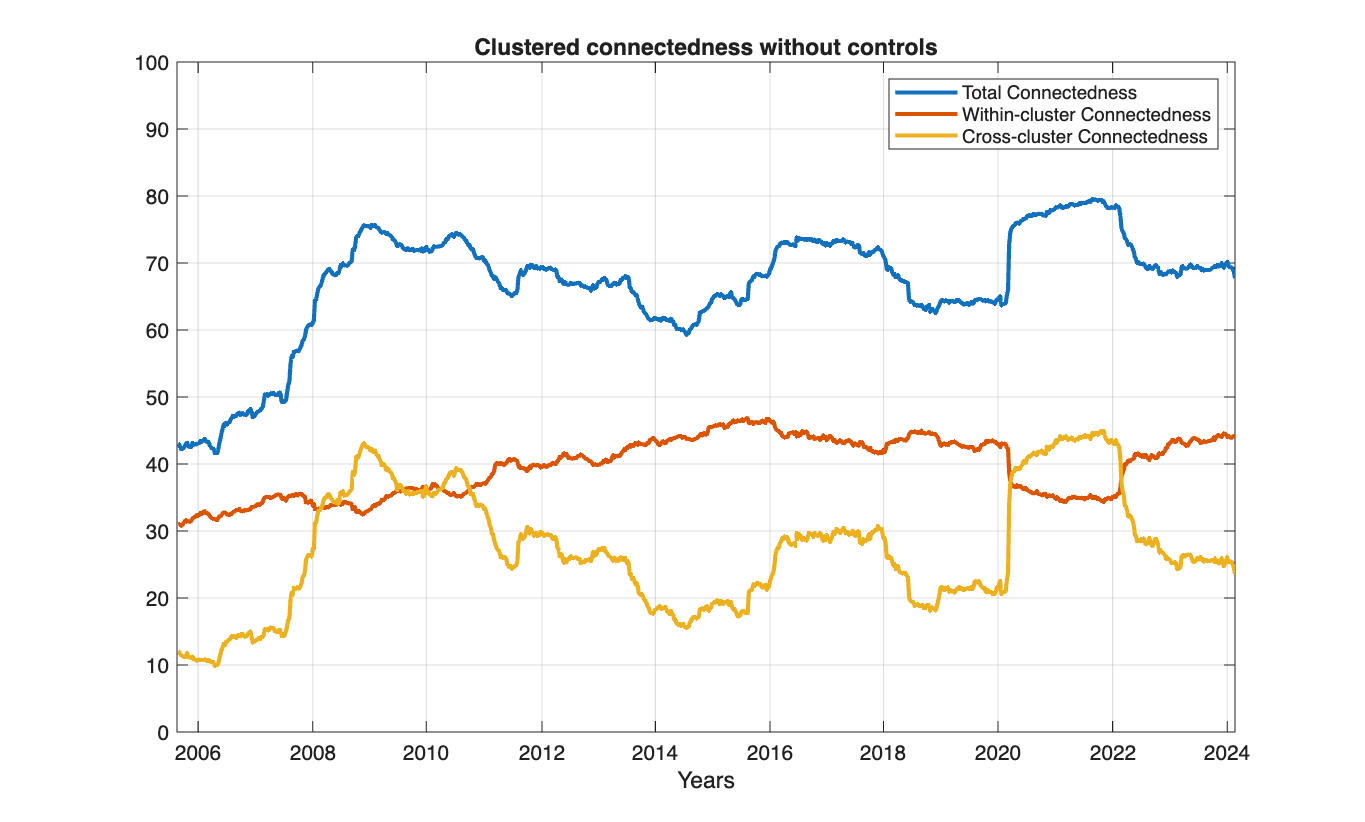}
			\caption{Without controls.}
			\label{fig:roll_idx_noctrl}
		\end{subfigure}
		\vspace{-0.4em}
		\begin{subfigure}{0.70\textwidth}
			\centering
			\includegraphics[width=\linewidth]{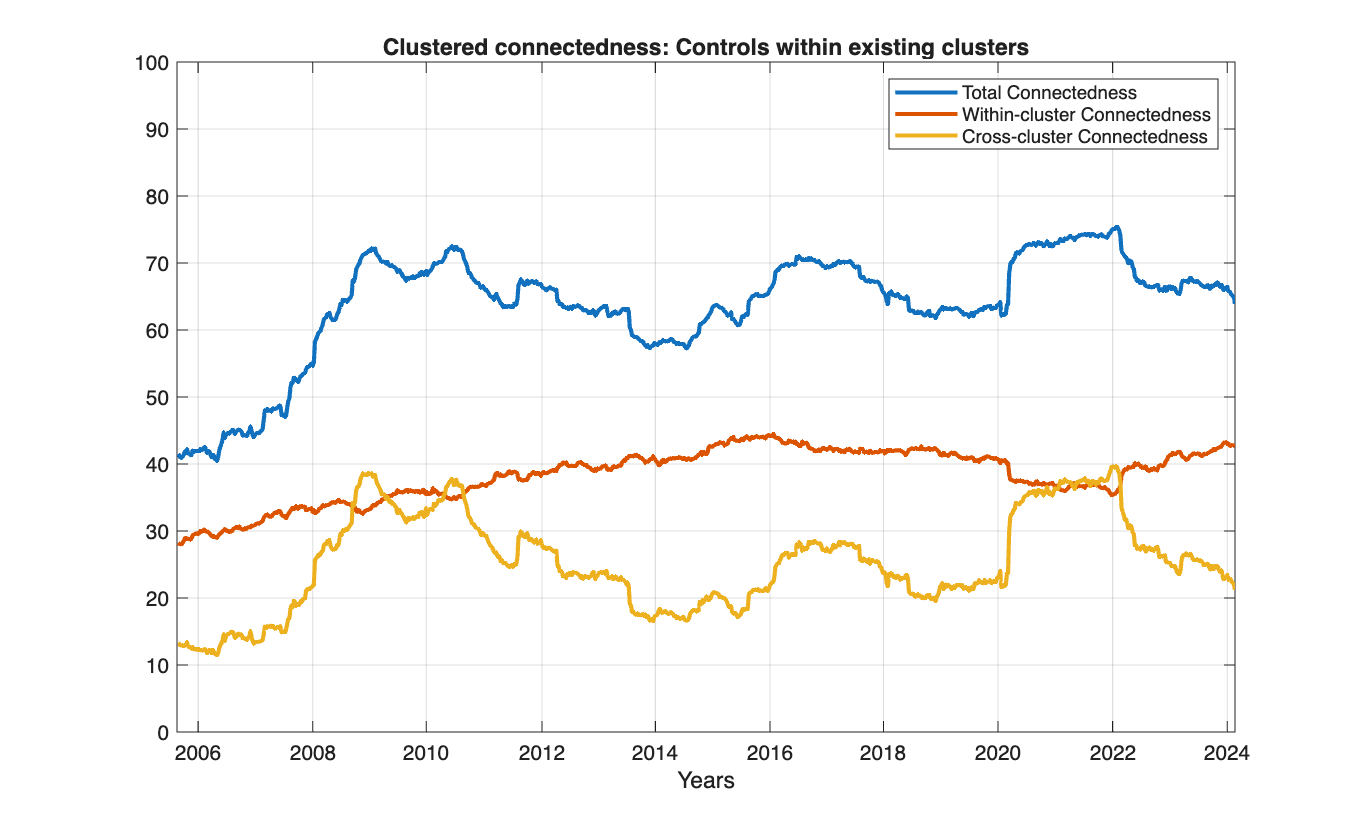}
			\caption{Controls within existing clusters.}
			\label{fig:roll_idx_ctrlwi}
		\end{subfigure}
		\vspace{-0.4em}
		\begin{subfigure}{0.70\textwidth}
			\centering
			\includegraphics[width=\linewidth]{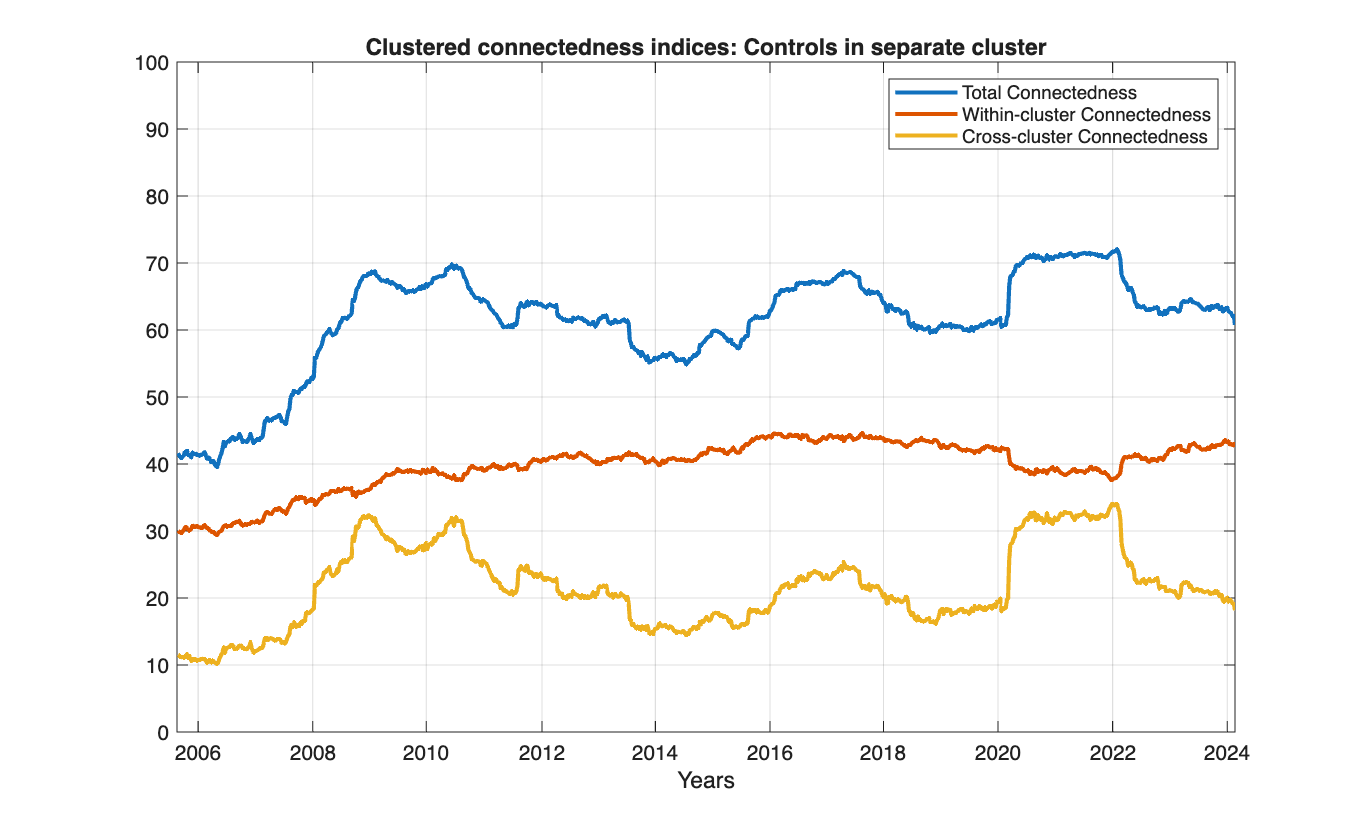}
			\caption{Controls in a separate cluster.}
			\label{fig:roll_idx_ctrlsep}
		\end{subfigure}
		\caption{Rolling-window system-wide total, within-group, and cross-group connectedness indices (two-year window) across alternative treatments of the control variables.}
		\label{fig:roll_idx}
	\end{figure}

	Connectedness is not constant over time but responds strongly to episodes of financial stress. To examine those dynamics, we implement a rolling-window version of our analysis. We re-estimate the VAR(2) by elastic net over rolling windows of 504 trading days (two years), advancing the window one day at a time, and compute the connectedness measures over all $5{,}040$ orderings at every fifth trading day. Figure~\ref{fig:roll_idx} reports the resulting system-wide total, within-group, and cross-group connectedness indices for the three specifications, as defined in Subsection~\ref{subsec:connectedness_measures}: without controls in Panel~\subref{fig:roll_idx_noctrl}, with controls within existing clusters in Panel~\subref{fig:roll_idx_ctrlwi}, and with controls in a separate cluster in Panel~\subref{fig:roll_idx_ctrlsep}. Across all three panels, total connectedness rises sharply during the global financial crisis of 2008--2009, remains elevated through the euro-area sovereign debt crisis of 2011--2012, and spikes again during the COVID-19 pandemic of 2020--2022.
	
	Two findings are particularly noteworthy. First, within-group connectedness is barely affected by the inclusion of controls: the within-group series (red) is nearly identical in level and shape across the three panels, drifting upward from roughly thirty to the low forties over the sample regardless of how the controls are treated. Second, cross-group connectedness is affected the most, as expected, since the controls account for common-factor variation that the cross-group component would otherwise partly reflect. The cross-group series (yellow) shifts down once the controls are introduced and is lowest when they form a separate cluster, with the crisis peaks falling from the mid-forties without controls to roughly forty with controls within the clusters and to the low thirties with a separate control cluster. Because total connectedness is the sum of the within-group and cross-group components, its response to alternative control specifications is driven almost entirely by the cross-group component. This reinforces a central conclusion of the paper: conditioning more explicitly on observed common-factor variation primarily changes measured system-wide cross-group connectedness while leaving system-wide within-group connectedness largely intact.
	
	\subsection{Ordering Sensitivity}
	\label{subsec:ordering_empirical}
	
	We next turn from the level and structure of measured connectedness to its robustness across admissible recursive identification orderings.
	
	\subsubsection{Identification Strategies}
	
	While within-cluster ordering is irrelevant for variance-decomposition-based measures, the ordering \emph{across} the seven clusters is not. With seven clusters we face $7! = 5{,}040$ possible orderings, and a natural question is whether our conclusions depend on any particular one. To address this we compute net connectedness for every admissible ordering and examine the resulting distributions.
	
\begin{figure}[tb!]
		\centering
		\includegraphics[width=0.9\textwidth]{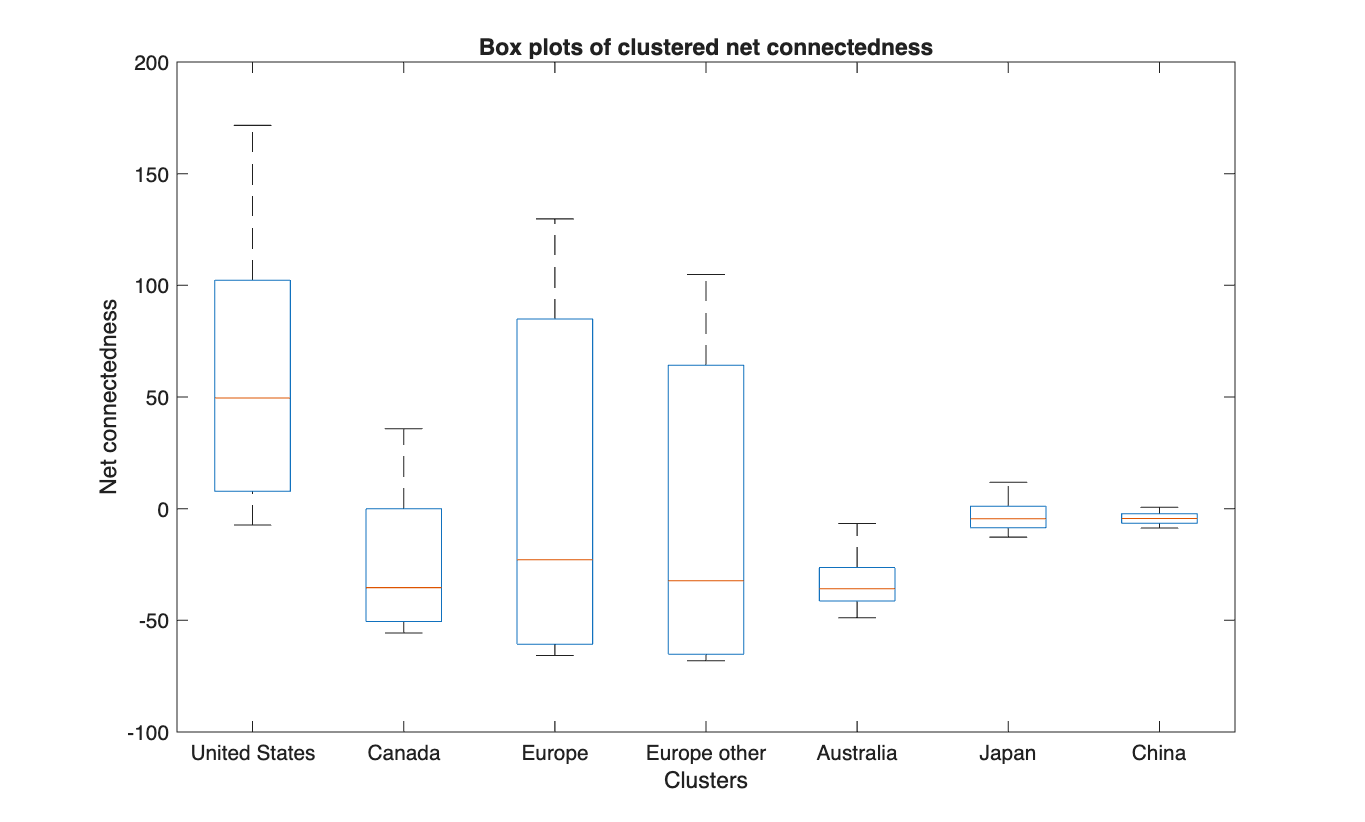}
		\caption{Box plots of clustered net connectedness across all $7!=5{,}040$ orderings (without controls).}
		\label{fig:boxplots}
	\end{figure}

	In Figure~\ref{fig:boxplots} we present box plots of cluster-level net connectedness across all $5{,}040$ orderings for the specification without controls. The U.S.\ cluster exhibits the widest dispersion, with its net connectedness ranging from roughly $-10$ to $+170$ percentage points depending on its position in the orthogonalization sequence. European clusters also show wide dispersion, reflecting their strong cross-cluster linkages. The Canadian and Australian distributions are concentrated below zero, with Australia's especially so; net-receiver status therefore holds for most orderings for Canada and appears especially robust for Australia. Japan and China exhibit narrow box plots, consistent with their comparatively weak measured linkages to the rest of the network.
	
	\subsubsection{Kernel Density Estimates}
	
\begin{figure}[tb!]
		\centering
		\includegraphics[width=0.85\textwidth]{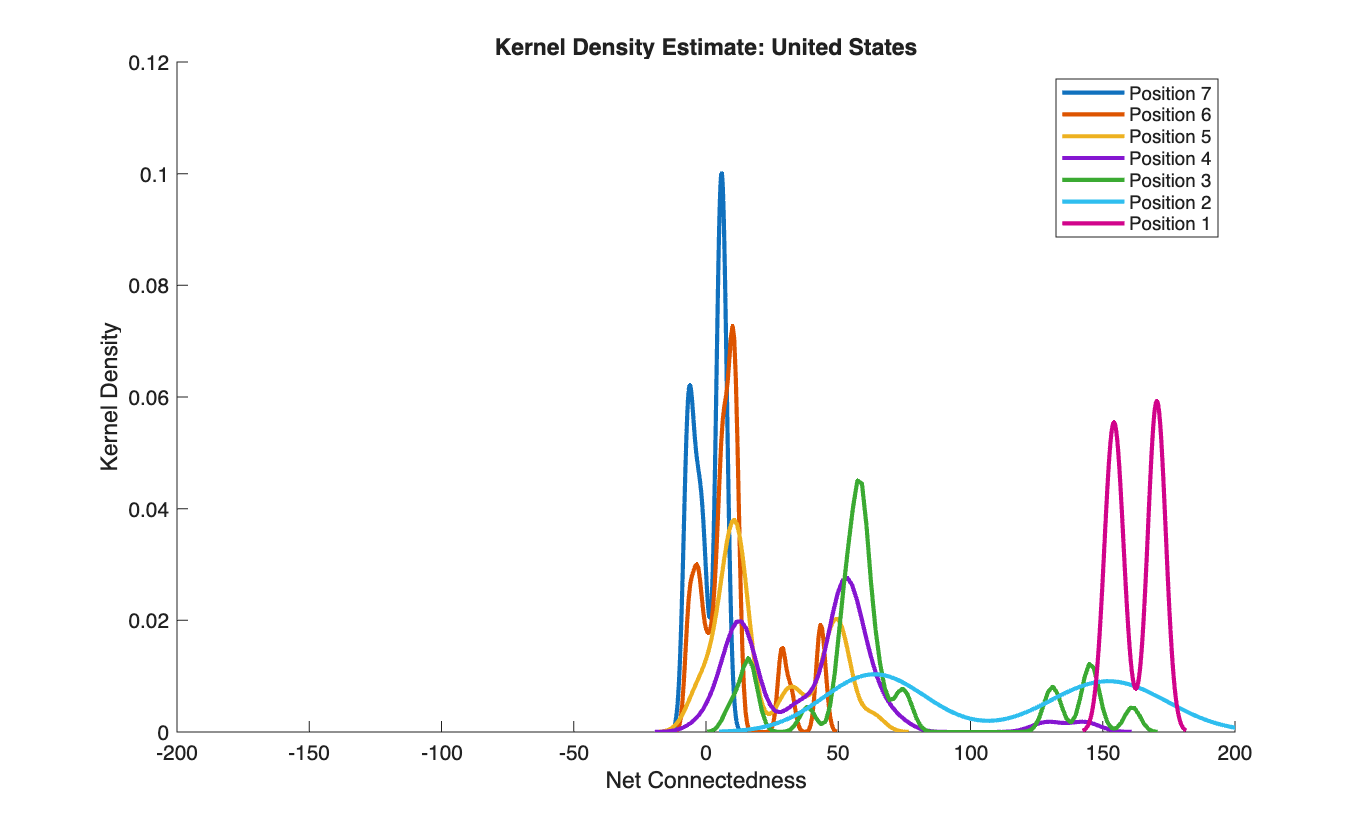}
		\caption{Kernel density estimates of U.S.\ net connectedness conditional on ordering position.}
		\label{fig:kde_usa}
	\end{figure}

\begin{figure}[tb!]
		\centering
		\includegraphics[width=0.95\textwidth]{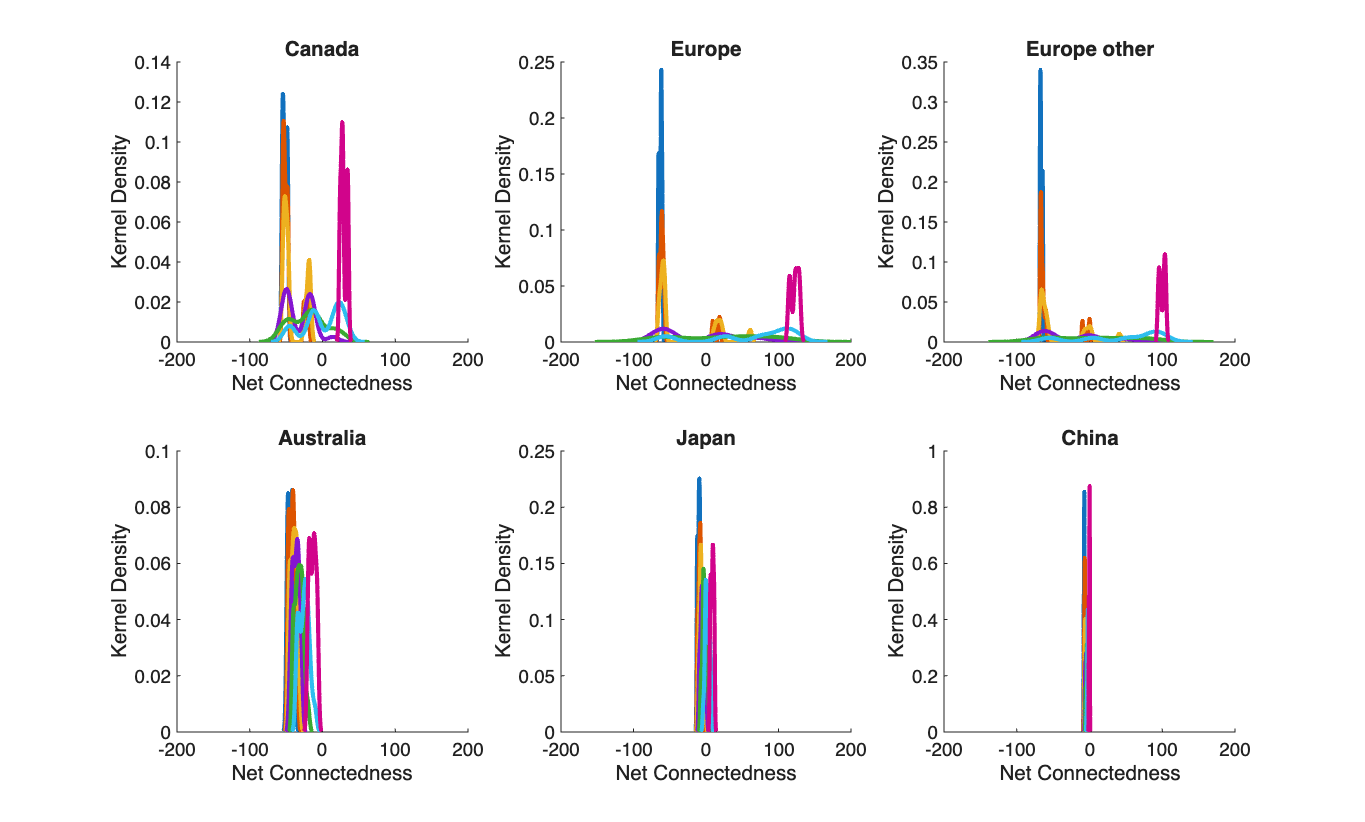}
		\caption{Kernel density estimates of net connectedness conditional on ordering position for the six non-U.S.\ clusters.}
		\label{fig:kde_6}
	\end{figure}

	To examine how a cluster's net connectedness depends on its position in the orthogonalization sequence, Figures~\ref{fig:kde_usa} and~\ref{fig:kde_6} report kernel density estimates of net connectedness conditional on position, separately for each cluster.
	
	In Figure~\ref{fig:kde_usa} we show the position-conditional densities for the United States. The distribution is clearly ordered: Position 1 (the U.S.\ orthogonalized first) generates the largest net connectedness values (density concentrated between roughly $+150$ and $+175$). As the U.S.\ moves later in the ordering, its net connectedness declines monotonically through Positions 2--7, with Position 7 (orthogonalized last) producing net connectedness near zero. This pattern shows that ordering matters quantitatively for the U.S.\ and motivates our strategy of averaging over all orderings.
	
	In Figure~\ref{fig:kde_6} we show analogous position-conditional densities for the six non-U.S.\ clusters (Canada, Europe, Europe Other, Australia, Japan, China). For Canada, Australia, Japan, and China, the dispersion across positions is markedly narrower than for the United States, particularly so for Japan and China. This is consistent with the relatively weak cross-cluster linkages of these economies: if a cluster has weak measured connections to the rest of the system, its ordering position has little influence on its net connectedness. The two European clusters are the exception, with dispersion comparable to that of the U.S.
	
	The two European clusters exhibit striking \emph{bipolar} distributions: when Europe is first in the ordering (Position 1), its net connectedness concentrates near $+120$ percentage points, whereas when Europe is last (Position 7), its net connectedness concentrates sharply near $-60$. The intermediate positions produce distributions that lie between these two extremes, generally favoring the late-ordered (negative) side. This bipolar pattern, particularly pronounced for EUO, where Position 7 generates an exceptionally tall spike near $-70$, is consistent with Europe's dense linkages to both the U.S.\ banking cluster and the rest of Europe, which make its measured net connectedness especially sensitive to how contemporaneous correlation with these clusters is allocated.
	
	Figure~\ref{fig:kde_6} also displays the density for Canada, which exhibits a similar but attenuated bipolar structure: Position 1 concentrates near $+30$, Position 7 near $-55$, with intermediate positions forming broader, flatter distributions and the negative side clearly dominating. Canada's close measured linkages to the U.S.\ cluster are consistent with this ordering sensitivity.
	
	In the case of Australia the dispersion across positions is narrower than for Canada or Europe but remains nontrivial: Position 1 concentrates near $-10$ and Position 7 near $-45$, a span of roughly $35$ percentage points. Notably, all of the position-conditional densities are concentrated below zero, so Australia remains a net receiver across the position-specific comparisons.
	
	Finally, for the Japanese and Chinese clusters, the position-conditional distributions are all concentrated within a narrow range near zero. Japan's densities span approximately $[-13, +10]$ and China's span approximately $[-8, 0]$, consistent with the comparatively weak measured linkages of these banking systems to the global network. In both cases, ordering sensitivity is economically small.
	
	\subsubsection{First-versus-Last Identification: A Robustness Measure}
	
	As an additional summary of ordering sensitivity, we use the paired first-versus-last diagnostic defined in Subsection~\ref{subsec:diagnostic}. For each cluster $c$ and each $\pi \in \Pi_{-c}$, we compute $\Delta_{c,\pi}^{(1,7)} = \mathit{Net}_{c,\pi}^{(1)} - \mathit{Net}_{c,\pi}^{(7)}$, where the relative ordering $\pi$ of the other six bank clusters is held fixed. We report the kernel density of $\Delta_{c,\pi}^{(1,7)}$ across $\pi$ for each cluster $c$ under three specifications: without controls, with controls placed within the existing clusters, and with controls placed in a separate cluster. As established there, if ordering did not matter this distribution would concentrate near zero; the farther from zero, and the wider the density, the more ordering sensitivity we observe.
	
\begin{figure}[tb!]
		\centering
		\includegraphics[width=0.85\textwidth]{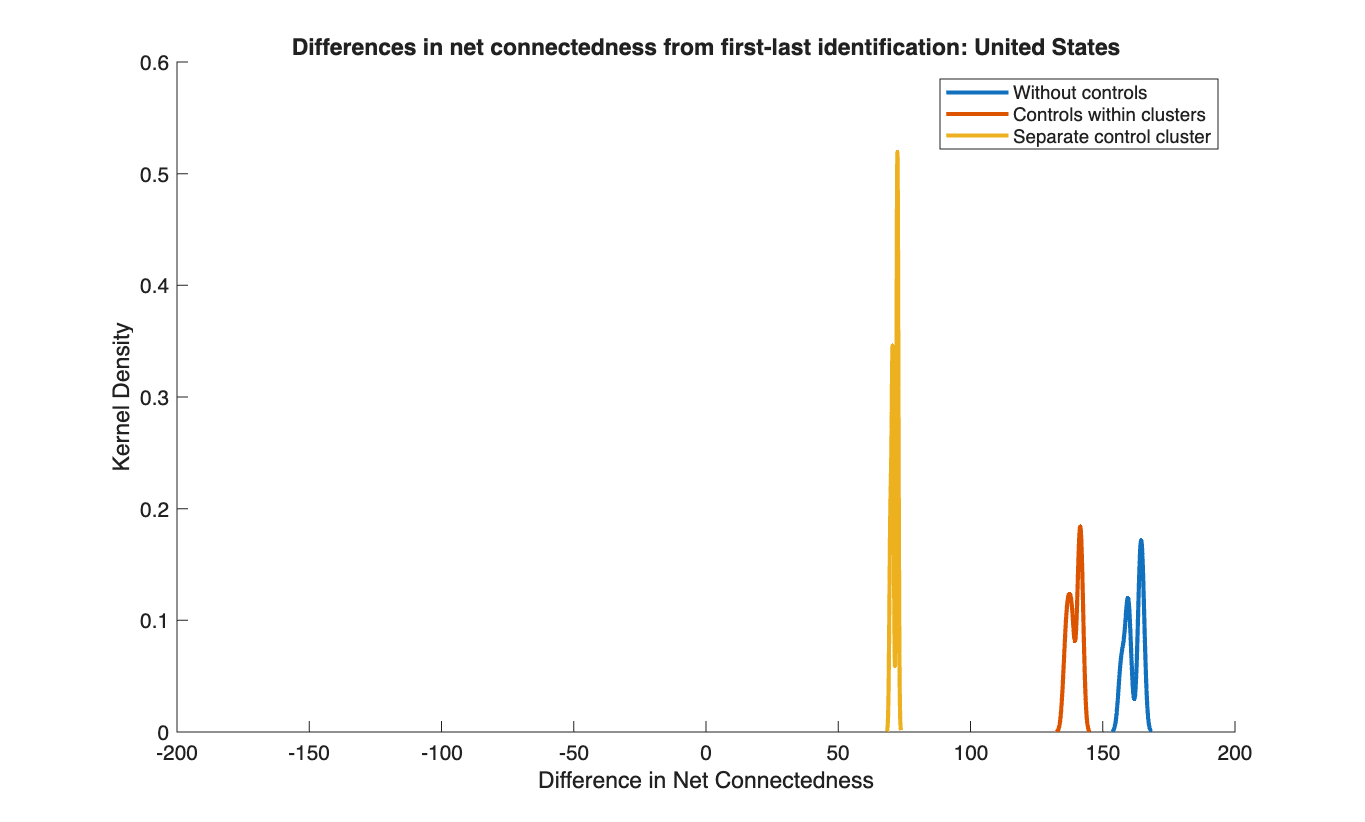}
		\caption{Distribution of the paired first-versus-last difference in U.S.\ net connectedness, without controls (blue), with controls within the existing clusters (red), and with controls in a separate cluster (yellow).}
		\label{fig:first_last_usa}
	\end{figure}

	Figure~\ref{fig:first_last_usa} displays the U.S.\ density. Without controls (blue), $\Delta_{\mathrm{USA},\pi}^{(1,7)}$ is tightly concentrated near $+160$, reflecting that, holding fixed the relative ordering of the other six clusters, U.S.\ net connectedness is typically about $160$ percentage points larger when the U.S.\ is orthogonalized first than when last. Placing the controls within the existing clusters (red) shifts the mode only modestly, to roughly $+140$. Placing the controls in a separate cluster (yellow) produces by far the largest improvement: the density collapses to a tall, narrow spike near $+72$, less than half the no-controls value and visibly tighter than either alternative. The ordering of the three specifications is a central empirical finding: a dedicated control cluster, retained in the system, delivers the most ordering-robust U.S.\ net connectedness.
	
\begin{figure}[tb!]
		\centering
		\includegraphics[width=0.95\textwidth]{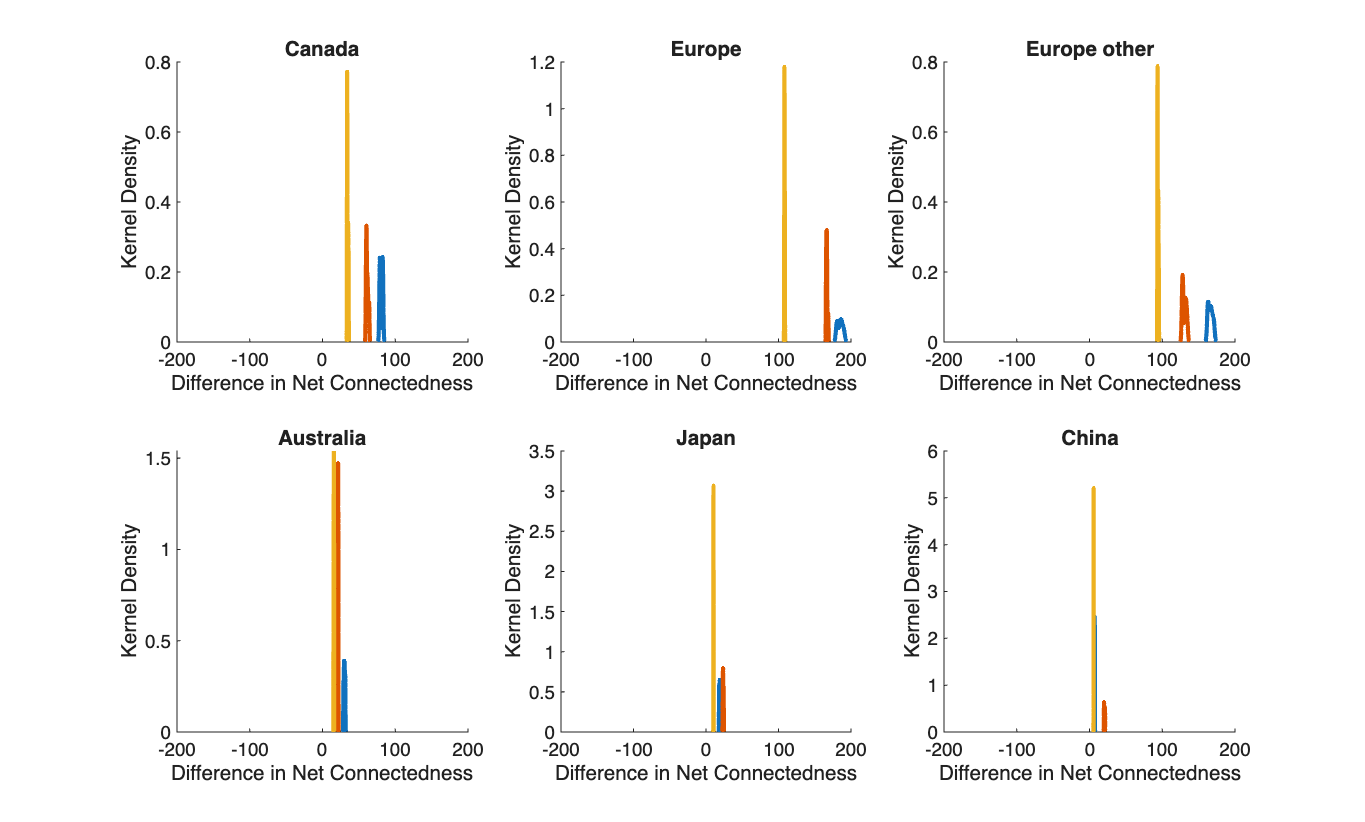}
		\caption{Distribution of the paired first-versus-last difference in net connectedness, six non-U.S.\ clusters, without controls, with controls within the existing clusters, and with controls in a separate cluster.}
		\label{fig:first_last_6}
	\end{figure}

	In Figure~\ref{fig:first_last_6} we show the corresponding distributions for the six non-U.S.\ clusters. For Europe and Europe Other the no-controls magnitudes match or exceed those of the U.S.\ (modes near $+185$ and $+160$), while the remaining clusters are smaller. The separate-control-cluster densities (yellow) are the closest to zero and the tightest for every cluster. The within-cluster densities (red) are intermediate for the larger clusters, but for Japan and China they sit slightly farther from zero than the no-controls densities, so embedding the controls in the clusters helps the most ordering-sensitive clusters while modestly increasing the difference for the least sensitive ones. For China, the difference is small without controls and with the separate control cluster, consistent with its comparatively weak measured linkages to the global banking network. Taken together, these distributions provide direct empirical evidence that the proposed control-variable framework substantially reduces ordering sensitivity. By assigning variation associated with observed common macro-financial factors to a dedicated control cluster, the remaining measured bank-to-bank connectedness becomes markedly less dependent on the ordering of the bank clusters.
	
	\subsubsection{Ordering Sensitivity Over Time}

	At each date $t$, we compute the paired first-versus-last differences $\Delta_{c,\pi,t}^{(1,7)}$ across all $\pi \in \Pi_{-c}$, where the subscript $t$ indicates that connectedness is estimated using the two-year rolling window ending at date $t$. We then plot the mean paired difference
	\begin{equation*}
		\overline{\Delta}_{c,t}^{(1,7)}
		=
		\mathbb{E}_{\pi}\!\left[\Delta_{c,\pi,t}^{(1,7)}\right]
	\end{equation*}
	over time. A flat line near zero indicates little average sensitivity to the first-versus-last change in ordering position at that date, while variation away from zero indicates periods of heightened ordering sensitivity. We report these mean paired differences across three specifications: without controls, with controls placed within their respective clusters, and with controls placed in a separate cluster.
	
\begin{figure}[tb!]
		\centering
		\includegraphics[width=0.95\textwidth]{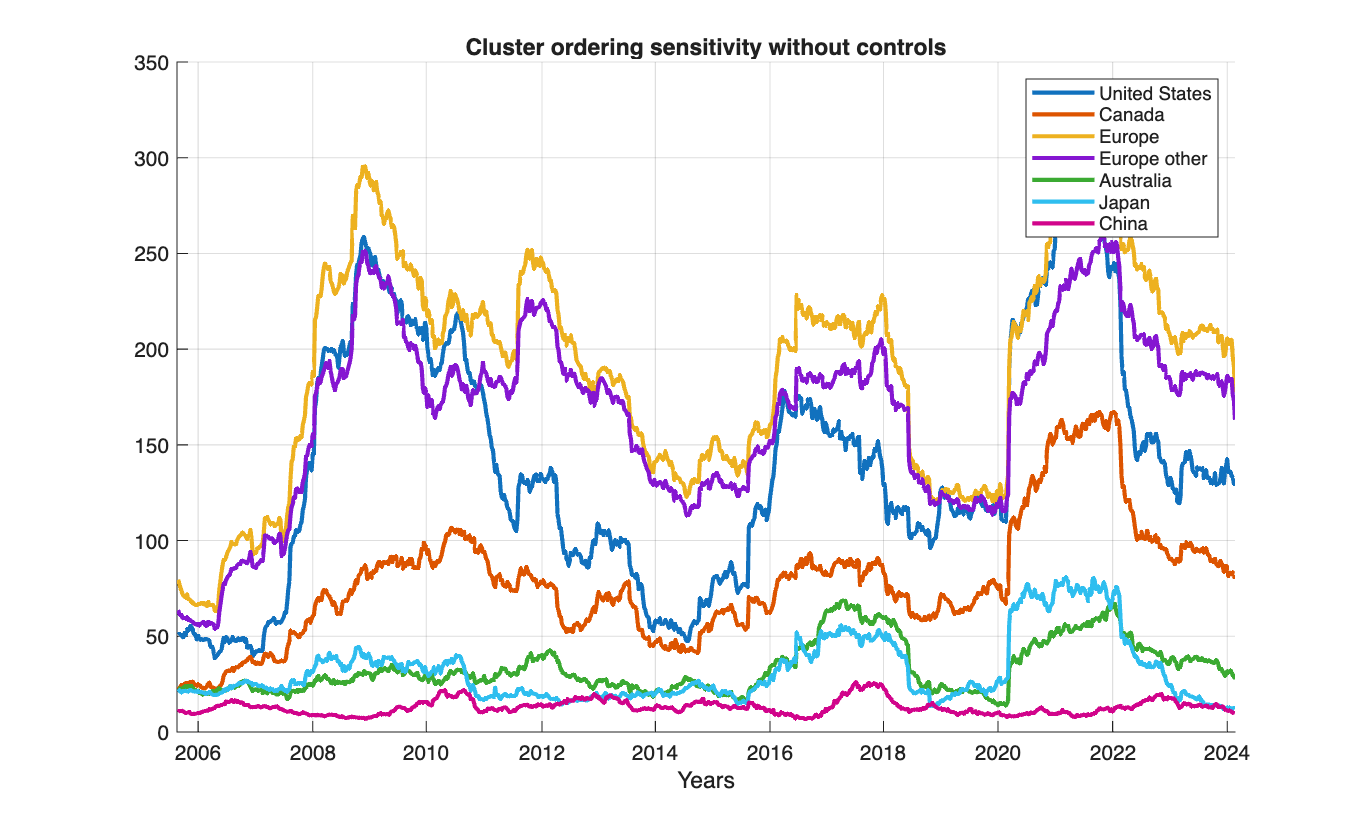}
		\caption{Rolling-window mean paired first-versus-last difference in net connectedness by cluster (two-year window), without controls.}
		\label{fig:roll_delta_noctrl}
	\end{figure}

\begin{figure}[tb!]
		\centering
		\includegraphics[width=0.95\textwidth]{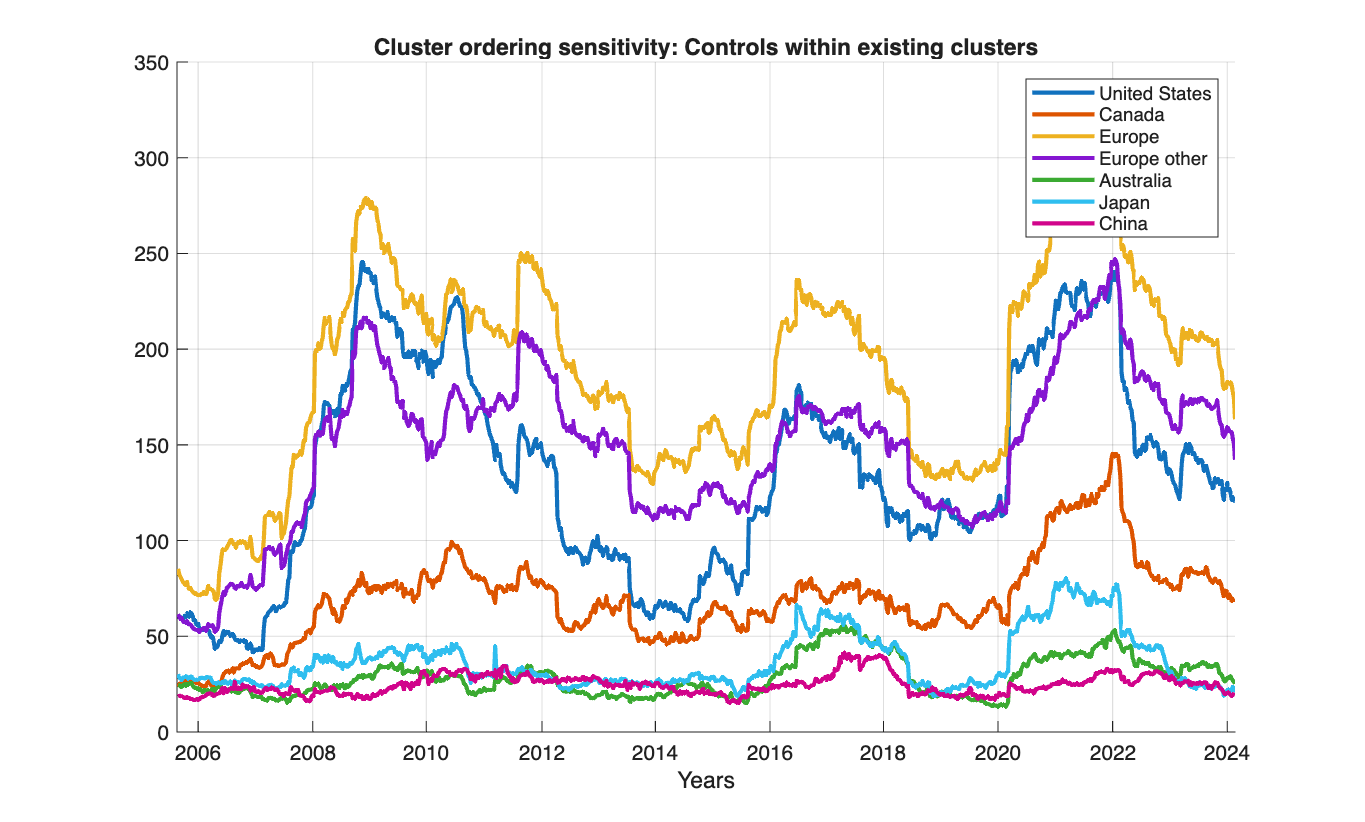}
		\caption{Rolling-window mean paired first-versus-last difference in net connectedness by cluster (two-year window), with controls within existing clusters.}
		\label{fig:roll_delta_ctrlwi}
	\end{figure}

\begin{figure}[tb!]
		\centering
		\includegraphics[width=0.95\textwidth]{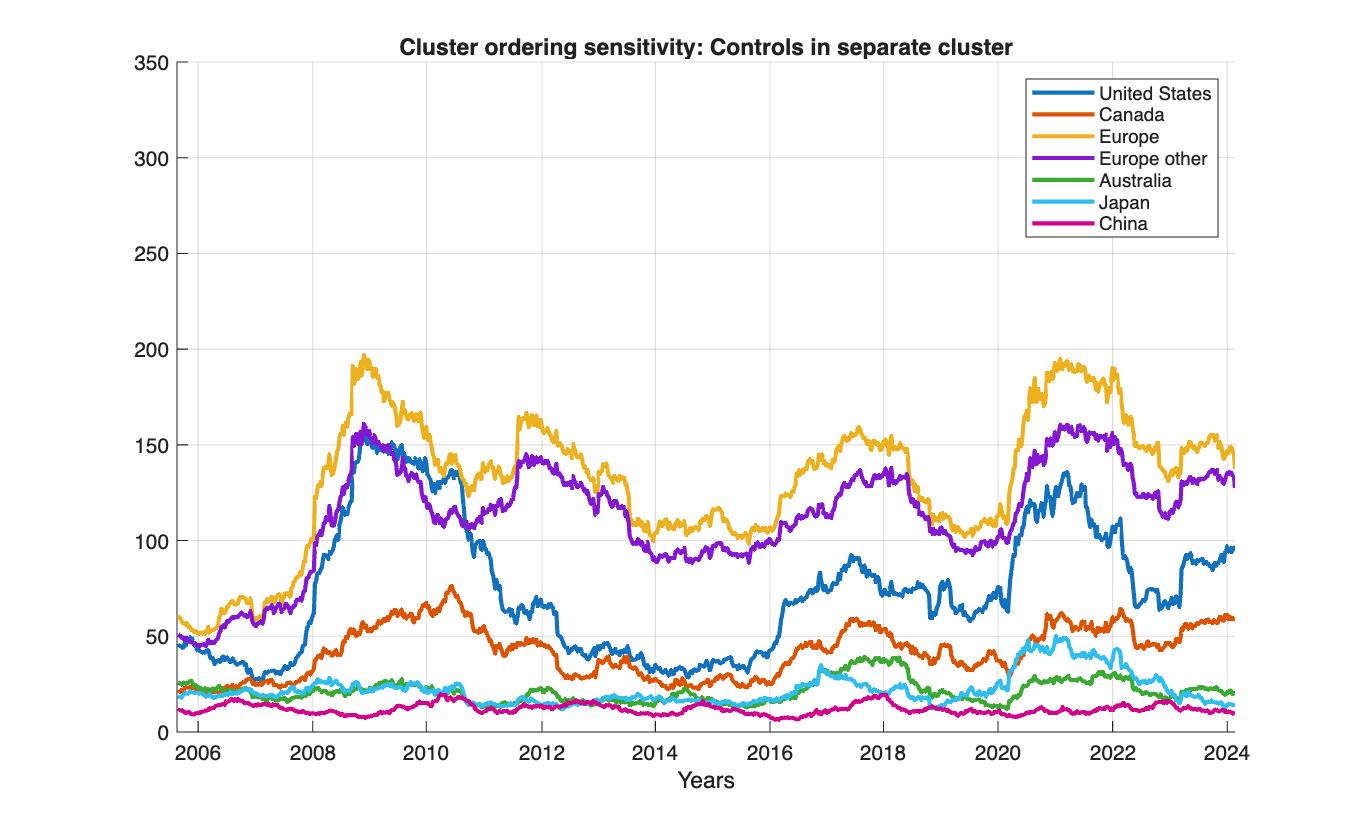}
		\caption{Rolling-window mean paired first-versus-last difference in net connectedness by cluster (two-year window), with controls in a separate cluster.}
		\label{fig:roll_delta_ctrlsep}
	\end{figure}

	In Figures~\ref{fig:roll_delta_noctrl}, \ref{fig:roll_delta_ctrlwi}, and~\ref{fig:roll_delta_ctrlsep} we display the rolling-window mean paired first-versus-last differences in net connectedness by cluster for the same three specifications. Clear peaks emerge around the global financial crisis of 2008--2009, the euro-area sovereign debt crisis of 2011--2012, the period of 2016--2018, and the COVID-19 pandemic of 2020--2022. In every specification the U.S., Europe, and Europe Other clusters display the largest swings, while Japan, China, and Australia remain consistently low.
	
	The comparison across specifications delivers the central message. Without controls (Figure~\ref{fig:roll_delta_noctrl}), the mean paired first-versus-last differences for the most connected clusters reach as high as roughly two hundred ninety-five percentage points at the crisis peaks. When we place the controls within their respective clusters (Figure~\ref{fig:roll_delta_ctrlwi}), the impact is small: the peaks compress only modestly, to roughly two hundred seventy-five percentage points, and the overall pattern is largely unchanged. When we instead place the controls in a separate cluster (Figure~\ref{fig:roll_delta_ctrlsep}), the mean paired first-versus-last difference for each region decreases substantially, with the largest peaks falling to roughly one hundred ninety-five percentage points. The secular level of ordering sensitivity is lower throughout the sample.
	
	The time paths in Figure~\ref{fig:roll_delta_noctrl} track the crisis history of the sample. During the global financial crisis of 2008--2009, ordering sensitivity rises steeply not only for the U.S.\ but also for EUR and EUO, a pattern consistent with the especially strong U.S.--European linkages evident elsewhere in the results. As the U.S.\ crisis subsided, the U.S.\ first-versus-last difference slid down, while the European differences remained elevated through the euro-area banking and sovereign debt crisis, receding only around 2014. All three rise again over 2016--2018, a period that combines renewed European banking stress (Italian nonperforming loans and profitability concerns under negative interest rates) with the Brexit referendum and the Federal Reserve's tightening cycle. The COVID-19 shock then produces a sharp common jump in the U.S., EUR, and EUO differences,  which the two-year window keeps elevated through early 2022. Among the remaining clusters, only Canada displays comparable time variation, most visibly around the COVID-19 shock; Australia, Japan, and China remain low and relatively flat throughout.

	The controls change the level of these paths more than their shape. Placing the controls within the existing clusters (Figure~\ref{fig:roll_delta_ctrlwi}) leaves the time-series behavior of the U.S., EUR, and EUO essentially unchanged. Placing them in a separate cluster (Figure~\ref{fig:roll_delta_ctrlsep}) lowers every path
substantially, most visibly for the U.S., EUR, and EUO, yet the peaks and troughs remain in the same places. Ordering sensitivity therefore tends to be largest during periods of broad common stress, and the control cluster accounts for much of the associated common variation, but the crisis chronology survives in what remains.

	These time-series patterns are consistent with the full-sample distribution plots in Figures~\ref{fig:kde_usa}--\ref{fig:first_last_6}. As we include more relevant controls in a separate cluster, the difference that ordering makes shrinks, because variation associated with observed common factors is increasingly assigned to the control cluster rather than allocated across the bank clusters according to their orthogonalization position. More broadly, these time-series results show that the reduction in ordering sensitivity is not confined to the full-sample averages. Rather, the proposed control-variable framework delivers more robust connectedness measures throughout the sample, including during periods of severe financial stress when ordering sensitivity is greatest.

	\subsection{Identification and Economic Interpretation}
	\label{subsec:interpretation}
	
	We now turn from ordering sensitivity to economic interpretation. The generalized approach and the three clustered approaches (without controls, with controls placed within the existing clusters, and with controls in a separate cluster) provide complementary views of the same banking system, allowing us to assess both how identification affects measured connectedness and which empirical features are robust to alternative treatments of common macro-financial factors.
	
	\subsubsection{Overview}
	
	The system-wide aggregate connectedness statistics in Table~\ref{tab:agg_conn} and the cluster-level net connectedness values in Tables~\ref{tab:clustered}, \ref{tab:clustered_ctrl}, and~\ref{tab:clustered_ctrl_sep} together describe how identification choices and the treatment of common factors reshape our view of the global banking network.
	Three broad patterns emerge. First, \emph{clustering reduces measured connectedness relative to the generalized approach}, and the reduction operates almost entirely through the cross-cluster component. In the full sample (Panel~A of Table~\ref{tab:agg_conn}), total connectedness falls from $76.49$ under the generalized approach to $67.34$ with controls in clusters and $65.38$ with a separate control cluster, while within-cluster connectedness barely moves (between $39.36$ and $41.35$). Cross-cluster connectedness, by contrast, drops from $37.13$ to $25.99$ and $25.04$. The gap reflects co-movement that remains embedded in the generalized measure but is treated differently under the clustered specifications.
	
	Second, \emph{the placement of the controls is what determines how much control-related common variation is separately attributed}. The two placements deliver similar aggregate levels: dispersing the controls within the existing clusters lowers the aggregates almost as much as the separate cluster does. Where they differ is in what is separately attributed. When the controls are gathered into a separate first cluster, the bank innovations are residualized with respect to that cluster before the bank clusters are orthogonalized. Control-related common variation is therefore attributed to the control cluster: the control cluster is by far the largest net transmitter ($+48.85$; Table~\ref{tab:clustered_ctrl_sep}) while receiving almost nothing in return. Under the maintained control-first identification, this specification provides the cleanest separation between variation associated with the observed controls and the remaining bank-cluster connectedness and, as shown in Subsection~\ref{subsec:ordering_empirical}, the lowest ordering sensitivity.
	
	Third, \emph{the U.S.\ net-transmitter role is especially sensitive to the treatment of the controls}. In Table~\ref{tab:us_evolution} we track U.S.\ cluster net connectedness across the three clustered specifications.
	
	\begin{table}[tb!]
		\centering
		\caption{U.S.\ Cluster Net Connectedness Across Specifications (Clustered Approach)}
		\label{tab:us_evolution}
		\smallskip
		\begin{tabular}{l rrr}
			\toprule
			& To & From & Net \\
			\midrule
			Without controls & 82.24 & 22.76 & $+59.48$ \\
			Controls in clusters & 62.04 & 16.03 & $+46.01$ \\
			Separate control cluster & 33.76 & 19.58 & $+14.19$ \\
			\bottomrule
		\end{tabular}
	\end{table}
	
	Without controls, the U.S.\ is the dominant net transmitter in the system ($+59.48$). When the controls are placed within the clusters, the U.S.\ net position weakens moderately ($+46.01$), but it is still the leading transmitter. When the controls are gathered into a separate cluster, variation associated with the observed common factors is attributed to the control cluster before the bank clusters are orthogonalized, and the net position of the U.S.\ attenuates substantially ($+14.19$). Under the maintained control-first identification, the economic interpretation is that a substantial share of the net transmission previously attributed to the U.S.\ is associated with common macro-financial factors (VIX, VSTOXX, VXJ, and VHSI) rather than with shocks attributed specifically to the U.S.\ banking cluster.  Yet in the separate-control specification the U.S.\ remains the only bank cluster with positive net connectedness. 
	
	\subsubsection{European Banks: Tight Integration, Distinct Net Roles}
	
	The European banking clusters (EUR and EUO) exhibit features that are remarkably stable across specifications.
	
	First, the bilateral spillovers between EUR and EUO are the strongest pairwise cluster linkages in every specification. The EUR-to-EUO flow is $24.71\%$ without controls, $24.20\%$ with controls in clusters, and $20.29\%$ with a separate control cluster; the EUO-to-EUR flow is $22.13\%$, $18.91\%$, and $16.97\%$ respectively. This bilateral channel survives the introduction of controls (even in the separate-cluster specification it remains the dominant cross-cluster linkage), consistent with strong institutional and financial integration rather than solely common-factor exposure: shared regulatory frameworks (CRD/CRR, SSM, SRB), cross-border interbank lending within the euro area, and common sovereign risk exposures.
	
	Second, the two European clusters play distinct roles. EUR is a modest net transmitter without controls ($+8.01$) and with controls in clusters ($+11.75$), and moves to near balance ($-2.90$) once the controls form a separate cluster. EUO, by contrast, is a net receiver in every specification ($-4.09$, $-8.34$, and $-13.83$), receiving heavily from EUR in each. The two clusters' large mutual flows and modest net positions indicate that European connectedness is dominated by dense two-way linkages rather than one-directional transmission.
	
	Third, the within-cluster co-movement among European banks is exceptionally dense. The own-cluster (diagonal block) shares for EUR and EUO are roughly $51$--$65\%$ across specifications, indicating that about half to two-thirds of European bank volatility variation is explained by co-movement with other banks in the same regional cluster. This feature motivates the block-Gram-Schmidt identification of the BDY approach: the strong within-cluster European co-movement makes it especially important to distinguish within-cluster dependence from identified cross-cluster transmission.
	
	\subsubsection{The Periphery: Canada, Australia, Japan, and China}
	
	The remaining four clusters play distinctive roles in the network. 
	
	\emph{Canada} is the cluster most exposed to U.S.\ banking shocks, with $24.30\%$ of its FEVD attributable to U.S.\ banks in the no-controls specification. It is a substantial net receiver in every specification ($-22.92$ without controls, $-15.96$ with controls in clusters, and $-19.59$ with a separate control cluster), consistent with its tight financial integration with the much larger U.S.\ system: Canadian banks absorb far more variance from U.S.\ banks than they transmit back.
	
	\emph{Australia} is one of the largest net receivers in the system ($-33.16$ without controls, $-21.33$ with controls in clusters, and $-19.68$ with a separate control cluster), absorbing substantial variance from the U.S.\ and European clusters while transmitting very little. Its own-cluster share ranges between $58\%$ and $76\%$ across specifications, and treating the controls explicitly absorbs part of its measured exposure.
	
	\emph{Japan} is a mild net receiver in all specifications ($-3.15$, $-3.04$, $-3.76$), with relatively low total connectedness to the Western banking clusters. Japanese banks' own-cluster share is $88$--$90\%$, consistent with the comparatively limited measured external connectedness of the Japanese banking system and the concentrated structure of Japanese banking (the ``megabank'' system of MUFG, SMFG, and Mizuho).
	
	\emph{China} stands out as the most insulated cluster in terms of measured connectedness, with own-cluster shares between $84\%$ and $92\%$ across specifications and To- and From-connectedness that are among the lowest of all the bank clusters. It is a mild net receiver in all three clustered cases ($-4.17$, $-9.08$, $-3.28$). Notably, China is also the cluster least exposed to the control cluster, i.e., only $1.36\%$ of Chinese FEVD is attributable to the controls in the separate-cluster specification, consistent with comparatively limited measured exposure to the observed global macro-financial controls. Taken together, these four clusters illustrate that the effects of the proposed identification strategy are heterogeneous: highly integrated banking systems are affected substantially by the treatment of common factors, whereas relatively insulated systems remain largely unchanged across specifications.
	
	\subsubsection{Summary of Findings}
	
	The comparative analysis yields five principal findings:
	
	\begin{enumerate}
		\item \textbf{Identification matters quantitatively.} Moving from the generalized to the clustered approach reduces system-wide total connectedness (from $76.49$ to $65.38$ in the full sample once the controls are placed in a separate cluster) almost entirely through a compression of the system-wide cross-cluster component (from $37.13$ to $25.04$). The corresponding within-cluster component is largely invariant to the identification strategy.
		
		\item \textbf{Common factors matter, and their placement matters for attribution and identification robustness.} Gathering the VIX, VSTOXX, VXJ, and VHSI into a separate control cluster makes that cluster the dominant net transmitter ($+48.85$) with almost no reception, consistent with its intended role as the first-ordered carrier of observed common-factor variation. Dispersing the same controls within the existing clusters delivers similar aggregate levels, but gathering them in a separate cluster explicitly attributes the common variation to the control cluster and, as shown above, delivers the large reduction in ordering sensitivity.
		
		\item \textbf{The U.S.\ net transmitter role is substantially affected by the treatment of common factors.} U.S.\ net connectedness falls from $+59.48$ without controls to $+46.01$ with controls in clusters and to $+14.19$ with a separate control cluster, a decline of roughly three quarters from the no-controls value. Under the maintained separate-control-cluster identification, this large reduction indicates that much of the net transmission previously attributed to the U.S.\ is associated with common macro-financial factor exposure, although the U.S.\ remains the leading bank-cluster net transmitter in every specification.
		
		\item \textbf{European banking integration is a robust empirical feature.} The EUR--EUO bilateral channel carries $17$--$25\%$ of each cluster's FEVD in every specification and remains the dominant cross-cluster linkage even after the controls are separated out, consistent with strong institutional interconnections rather than solely common-factor exposure.

		\item \textbf{China is comparatively insulated in measured connectedness.} Chinese banks have own-cluster shares between $84\%$ and $92\%$ and act as mild net receivers in every clustered specification ($-4.17$, $-9.08$, $-3.28$). China is also the least exposed of all clusters to the control cluster, consistent with comparatively limited measured exposure to the observed global macro-financial controls.
	\end{enumerate}
	
	\noindent Collectively, these findings show that the proposed clustered framework with a dedicated control cluster does more than reduce measured connectedness. It separates variation associated with observed common macro-financial controls from measured bank-to-bank connectedness, yielding measures that are both more economically interpretable and more robust to alternative admissible orderings. Viewed together, Subsections~\ref{subsec:banking_connectedness}--\ref{subsec:interpretation} show how the measured connectedness structure changes when observed common macro-financial variation is treated explicitly rather than being left embedded in the banking clusters.

	\section{Concluding Remarks}
	\label{sec:conclusion}
	
	This paper extends the BDY clustered connectedness framework \citep{BuchwalterDieboldYilmaz2026} in two complementary directions. First, we develop an ordering-sensitivity diagnostic based on all admissible bank-cluster orderings, with paired first-versus-last comparisons that hold fixed the relative ordering of all other clusters. This isolates the effect of a cluster's position and provides a transparent cluster-level assessment of identification robustness. Second, we introduce a dedicated control-variable cluster for observed common macro-financial factors without increasing the number of bank-cluster orderings. The control cluster is fixed first and the bank innovations are residualized with respect to it, yielding the corresponding Schur-complement covariance matrix, after which the bank clusters are permuted as before. Under the maintained recursive identification in which control-cluster innovations are contemporaneously exogenous to bank-cluster innovations, the remaining cross-cluster connectedness among the bank clusters can be interpreted as bank-to-bank transmission net of those observed common-factor shocks. Without that structural interpretation, it describes bank-cluster connectedness after partialing out contemporaneous linear variation associated with the observed controls.
	
	We apply the framework to seventy-one global banks in seven regional clusters over 2003--2024, using the VIX, VSTOXX, VXJ, and VHSI as full-sample controls. Several substantive findings emerge. Moving from generalized to clustered identification compresses the system-wide cross-group component of measured connectedness while leaving the corresponding within-group component largely unchanged. The treatment of the controls then determines how observed common-factor variation is allocated. Gathering the controls into a separate first cluster makes that cluster the dominant net transmitter ($+48.85$) and reduces U.S.\ net connectedness from $+59.48$ without controls to $+14.19$, although the U.S.\ remains the leading bank-cluster net transmitter. Under the control-first identification, this large reduction indicates that a substantial portion of the transmission previously attributed to the U.S.\ is associated with variation captured by the observed common macro-financial controls. By contrast, the EUR--EUO bilateral channel remains the dominant bank-to-bank cross-cluster linkage across specifications, while Chinese banks remain comparatively insulated from the global system.
	
	The ordering-sensitivity analysis reinforces this message. The distribution of net connectedness across the $5{,}040$ admissible cluster orderings is wide for the U.S.\ and the two European clusters, which have the densest cross-cluster linkages, and narrow for Japan and China. Placing the controls in a separate cluster delivers the most ordering-robust measures for every bank cluster. Placing the controls within the existing clusters provides modest improvement for the most ordering-sensitive clusters, whereas for Japan and China---where ordering sensitivity is small to begin with---it can slightly increase the first-versus-last difference relative to the no-controls case. For the U.S., the paired first-versus-last difference, computed while holding fixed the relative ordering of the other clusters, falls to less than half of its no-controls value under the separate-control-cluster specification. More broadly, the dedicated control cluster substantially reduces the dependence of measured bank-cluster connectedness on the subsequent recursive ordering.
	
	The framework opens several avenues for future research. Most immediately, the ordering problem itself remains to be addressed more fully. In this paper we diagnose ordering sensitivity, average over admissible orderings, and reduce it with controls. In ongoing work, we develop an order-invariant clustered identification. Building on \cite{FrancisHansenTong2026}, we choose, among all identification matrices that impose orthogonality across clusters while preserving correlation within them, the one that maximizes the correlation between structural and reduced-form shocks. The resulting identification is order-invariant and has a simple closed form. Other natural extensions include frequency-domain decompositions along the lines of \cite{BarunikKrehlik2018} and formal break-point detection in our rolling-window analysis.
	    
	\bibliographystyle{Diebold}
	\bibliography{current_paper_references_phase2}
	
\end{document}